\documentclass{birkjour_t2}

\pdfoutput=1

\usepackage{amsmath}
\usepackage{amssymb}
\usepackage{cite}
\usepackage{pgfplots}
\pgfplotsset{compat=1.14}
\usepackage{hyperref}
\usepackage[linesnumbered,ruled,vlined]{algorithm2e}

\usepackage{graphicx}
\usepackage{caption}
\usepackage{subcaption}

\usetikzlibrary{calc}

\renewcommand{\figurename}{Fig.}
\newcommand{\Sref}[1] {Section~\ref{#1}}
\newcommand{\sref}[1] {section~\ref{#1}}
\newcommand{\fref}[1] {\figurename~\ref{#1}}
\newcommand{\Fref}[1] {Figure~\ref{#1}}

\begin{document}

\title{TiGL -- An Open Source Computational Geometry Library for Parametric Aircraft Design}

\author{Martin Siggel \and Jan  Kleinert \and Tobias Stollenwerk}
\address{%
German Aerospace Center (DLR)\\
Simulation and Software Technology\\
Linder Hoehe\\
Cologne, 51147\\
Germany}

\email{martin.siggel@dlr.de}

\author{Reinhold Maierl}
\address{%
Airbus Defence and Space\\
Structure Analysis and Optimization\\
Rechliner Str.\\
Manching, 85077\\
Germany}

\email{reinhold.maierl@airbus.com}

\date{Received: date / Accepted: date}

\begin{abstract}
This paper introduces the software TiGL:
TiGL is an open source high-fidelity geometry modeler that is used in the conceptual and preliminary aircraft and helicopter design phase.
It creates full three-dimensional models of aircraft from their parametric CPACS description.
Due to its parametric nature, it is typically used for aircraft design analysis and optimization.
First, we present the use-case and architecture of TiGL.
Then, we discuss it's geometry module, which is used to generate the B-spline based surfaces of the aircraft.
The backbone of TiGL is its surface generator for curve network interpolation, based on Gordon surfaces.
One major part of this paper explains the mathematical foundation of Gordon surfaces on B-splines
and how we achieve the required curve network compatibility.
Finally, TiGL's aircraft component module is introduced, which is used to create the external and internal parts of
aircraft, such as wings, flaps, fuselages, engines or structural elements.

\end{abstract}

\maketitle

\section{Introduction}
Optimizing airplane designs often requires a large consortium of engineers from many different fields to work together.
Every group of engineers works with a specific set of simulation tools, but almost all tools require information about the current design's geometry.
The TiGL Geometry Library (TiGL) \cite{tigl_github} generates three-dimensional airplane geometries from a standardized parametric description. It is a piece of software developed mainly at the German Aerospace Center (DLR), in cooperation with Airbus Defense and Space and RISC Software GmbH.
These geometries include the outer shape exposed to the surrounding airfield as well as the inner structure of the fuselage and wings that provides the necessary stability.

\begin{figure}[ht]
	\centering
\begin{tikzpicture}[scale=1.0]
\node[inner sep=0pt] (russell) at (0,0)
 {
   \includegraphics[height=5cm]{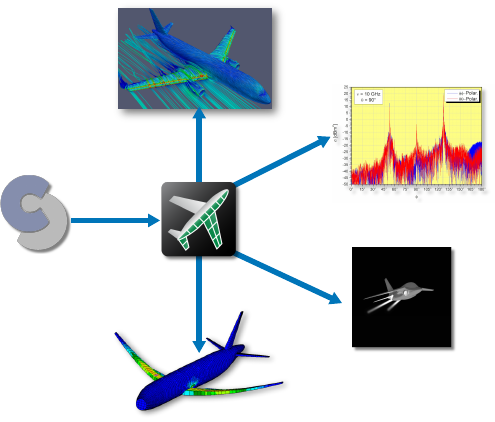}
 };
 \draw (-2.55, -0.8) node {CPACS};
 \draw (2.0, 1.7) node {Radar signature};
 \draw (2.2, -1.9) node {Infrared signature};
 \draw (-0.5, -2.7) node {Structure und Aeroelastics};
 \draw (-0.5, 2.7) node {Aerodynamics / CFD};
\end{tikzpicture}
\caption{TiGL is used as the central geometry pre-processor for many analysis tools inside and outside of the DLR.}
\label{fig:tigl_use}
\end{figure}
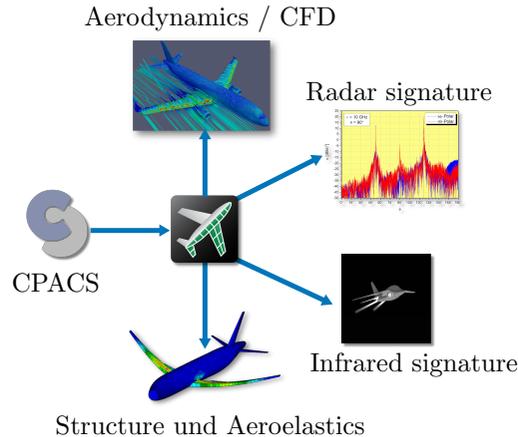

The \emph{Common Parametric Aircraft Configuration Schema} (CPACS)  is an exchange format for describing airplane design in form of an XML-file \cite{cpacs, cpacs_github}.
Among other things, it contains a parametric description of the aircraft geometry that is interpreted by TiGL.
TiGL offers the functionality to export these CPACS geometries to standard CAD formats such as IGES, STEP, VTK as well as functions to query points and curves on the airplanes surface.
TiGL uses the \emph{OpenCASCADE} CAD kernel \cite{opencascade} to model the geometries based on B-spline surfaces. Additional geometric modeling features are included on top of OpenCASCADE, such as specialized curve interpolation and approximation functions, surface skinning algorithms and a newly implemented algorithm to interpolate curve networks based on Gordon surfaces.
The library also provides interfaces to many common programming languages such as C, C++, Python, Java and MATLAB and comes with a graphical user interface to visualize a CPACS configuration.

	
TiGL is not the only freely available parametric geometry modeler for conceptual aircraft design.
While it is not the intention of this paper to compile a complete list, a few of these tools and publications deserve to be mentioned.
\emph{OpenVSP} \cite{openvsp} is a parametric aircraft design tool for aircraft developed by NASA.
Haimes and Drela published on the feasibility of conceptual aircraft geometry design for
high fidelity by using a bottom-up approach \cite{haimes}.
\emph{GeoMACH}~\cite{geomach} is a mutli-disciplinary analysis and optimization (MDAO) tool for geometric aircraft design, which supports a large number of design variables by providing also derivatives of the geometry with respect to the design variables.
\emph{Caesiom}~\cite{caesiom} is a design framework based on MATLAB that includes a parametric geometry modeler, but also simplified physical simulation tools for aerodynamics, structures, propulsion and flight control. 
\emph{SUMO}~\cite{sumo} is a surface modeler specifically designed for conceptual aircraft design that comes with a mesh generator and a post-processing tool.
Finally, \emph{JPAD}~\cite{jpad} is an analysis tool suite including \emph{JPADCAD}, an OpenCASCADE-based geometry modeler for aircraft.

This paper is organized as follows. 
Sections~\ref{sect:cpacs}~and~\ref{sect:optimization}  will introduce the idea behind CPACS and what role CPACS and TiGL play in multi-disciplinary optimization.
\Sref{sect:architecture} gives an overview of the software architecture and design.
\Sref{sect:geometry} describes TiGL's backbone, the geometry module. A special focus is given to the curve network interpolation algorithm used in TiGL to generate interpolating surfaces form a network of profile and guide curves.
The CPACS description and TiGL implementation for specific aircraft components such as wings, the airplane fuselage, control surfaces etc. are discussed in \Sref{sect:aircraftComponentModeling}.
Finally, the paper closes with a summary and outlook in \Sref{sect:conclusion}.

\subsection{CPACS Parametrization} \label{sect:cpacs}
The Common Parametric Aircraft Configuration Schema (CPACS) is a data model that contains parametric desciptions of aircraft configurations, as well as missions, airports, fleets and more \cite{cpacs}.
Its development started in 2005 at the German Aerospace Center, when there was an increasing need for a common aircraft model description that can be used in Multi-Disciplinary Optimization (MDO) applications.
It is specifically designed for collaborative design in a heterogenous environment of engineers from different fields. 
Engineers can use it to exchange information on their models and tools. 
Next to the model description, process information is stored so that CPACS can be used to setup interdisciplinary workflows.

CPACS it based on a schema definition (XSD) for XML and as such has a hierarchical structure.
On the highest level, the XSD description contains a header with meta-information about the CPACS file, such as a description, creation date, and CPACS version. 
Next to the header, there are elements for airlines, airports, flights, mission definitions, studies, tool specific information as well as vehicles. 
The latter element contains descriptions of airplanes or rotorcrafts, engines, fuels, materials, as well as guide and profile curves used for the geometric modeling of components.
TiGL uses the parametric description from these elements to construct the aircraft geometry.
\Sref{sect:aircraftComponentModeling} contains details on the parametric description of specific components as well as its interpretation in TiGL.
%
%
%
%

In late July 2018, CPACS 3.0 was released \cite{cpacs_github}. 
The upcoming release of TiGL 3.0 is tightly coupled to some of the major changes introduced in the new CPACS version. 
Revised definitions for \emph{``component segment''} coordinate systems or a new simplified definition of guide curve points are two reasons, why TiGL 3.0 is designed not to be backwards-compatible.
This means that TiGL 3 will not be able to read CPACS 2 files.
This is less error-prone and increases robustness of the code. 
To make the transition from CPACS 2 to CPACS 3 easier, a converter tool called \emph{"cpacs2to3"} is now being developed by the community \cite{cpacs2to3_github}. 
\subsection{Optimization} \label{sect:optimization}
\begin{figure}[ht]
\centering
\begin{subfigure}[t]{0.37\textwidth}
\includegraphics[width=\textwidth]{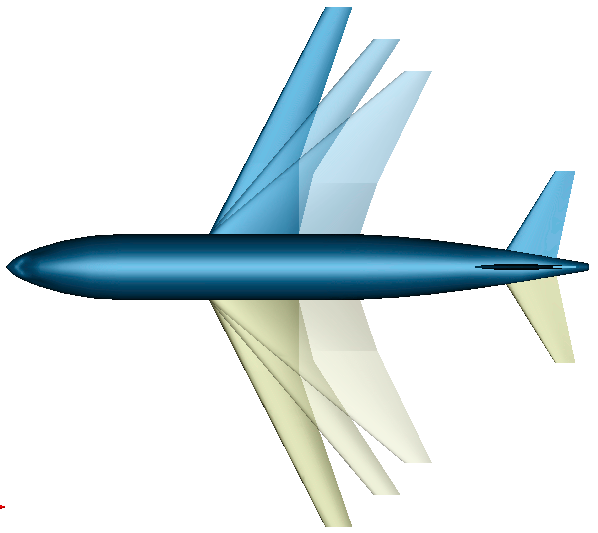}
\end{subfigure}
\qquad
\begin{subfigure}[t]{0.37\textwidth}
\includegraphics[width=\textwidth]{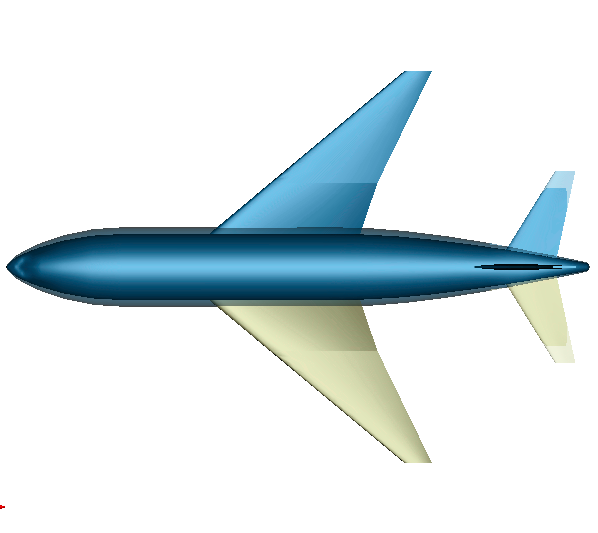}
\end{subfigure}
\caption{The shape of wings, fuselage and tailplane can smoothly be varied with just a few parameters.}
\label{fig:optimization}
\end{figure}

Both CPACS and the TiGL geometry library were designed specifically for use in mutli-disciplinary optimization (MDO).
The parametric description of CPACS enables users to directly control the configuration of an aircraft with just a small selection of parameters, see \fref{fig:optimization}.
With the help of TiGL, slight modifications of aircraft geometries can be created in an automated workflow.
CPACS and TiGL are currently being developed and constantly extentended in research projects related to MDO \cite{digitalx, mephisto1, victoria}.
Some ideas to increase the optimization capabilities in the future are 
\begin{itemize}
 \item to include automatic differentiation in TiGL's geometry module, 
 \item to automatically generate CFD meshes by including an open source mesh library, and
 \item to track mesh deformations through geometry changes and thus provide shape gradients with respect to parameters to an external optimization tool.
\end{itemize}

\section{Software Architecture} \label{sect:architecture}
As TiGL is open source software, all its dependencies
are open source as well.
In that sense, it can be used in its full functionality
without any commercial license by the users.
TiGL is mainly based on the TiXI XML library \cite{tixi_github}
to read or write the CPACS data sets.
TiGL heavily relies on the OpenCASCADE Technology CAD kernel \cite{opencascade},
which is used for the geometric and topological modeling,
for CAD data exports, to create solids via constructive solid geometry (CSG), 
and even for visualization.

Internally, TiGL contains multiple modules that are used for
the several different aspects of the software.
The geometry module includes all operations required to build
the curves and surfaces that finally resemble the aircraft shape.
These operations are all based on B-splines and NURBS (Non-Uniform Rational B-Splines).
In addition, it contains an extension to OpenCASCADE's
boundary representation (BREP) of shapes that adds metadata to the shapes.
These metadata contain information about the shape modification 
history and the names of the shape and its faces.
Since this information is preserved during CAD file exports, it can be used by external mesh generators,
e.g. to create boundary layers around the wing surface.

\begin{figure}[tb]
\centering
\subcaptionbox{\label{fig:architecture}}
	{\includegraphics[height=5cm]{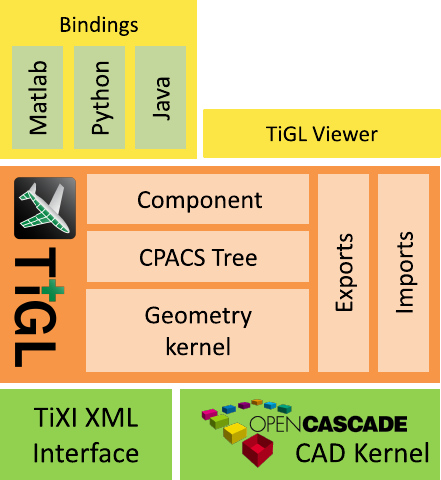}}
	\qquad
	\subcaptionbox{\label{fig:tiglviewer}}
	{\includegraphics[height=5cm]{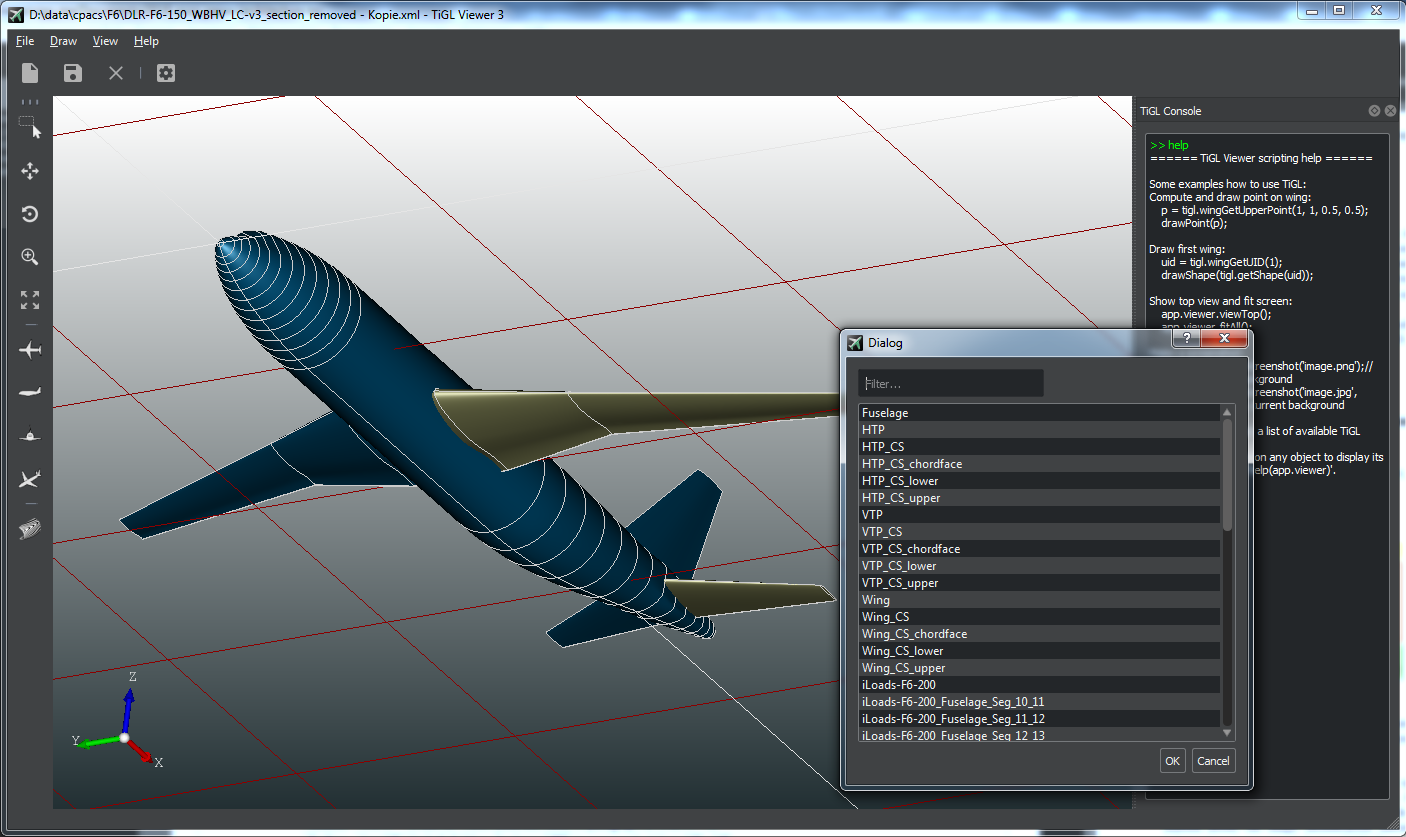}}
\caption{TiGL's system architecture (A) and a screenshot of the TiGL Viewer (B), which
is used to display CPACS geometries.}
\end{figure}

The CPACS tree module is an object hierarchy of the CPACS standard.
Each node in the CPACS tree is mapped to a C++ object that contains all of its
attributes and sub-nodes as child objects.
The code for these classes is automatically generated from
the CPACS XML schema (see Section~\ref{sec:auto_codegen}).

The component module implements the modeling of the all aircraft components and the
wing and fuselage structure, including spars, ribs, frames,
beams and much more (see \Sref{sect:aircraftComponentModeling}).
The export and import module are the interface to other
analysis tools.
TiGL can export the aircraft geometries to 
CAD-based as well as triangulated file formats.
The former includes standard CAD exchanges formats such as
IGES, STEP (ISO 10303) and OpenCASCADE's internal format BREP.
They are mainly used in combination with external mesh generation software.
The latter includes STL (stereolithography),
VTK polydata (to be used e.g. in ParaView \cite{paraview}),
and COLLADA	(COLLAborative Design Activity) to support
3D rendering of the geometries.

Although TiGL is written in C++, it 
provides bindings to 
C, MATLAB, Python, and JAVA (see \Sref{sec:bindings}).
In addition to the library, the software package comes
with \emph{TiGL Viewer}, a viewer for CPACS and 
other CAD files.
It uses TiGL for modeling and provides a 3D OpenGL
based view of the geometries.
In addition to the pure visualization, TiGL Viewer
includes a scripting console that can be used for
small automation tasks e.g. to create a four-sided view of the aircraft.
A screenshot of TiGL Viewer is shown in \fref{fig:tiglviewer}.

\subsection{Automatic Code Generation}
\label{sec:auto_codegen}
For every relevant CPACS entity, there must be a corresponding representation in TiGL.
This is achieved by automatic code generation from the CPACS schema 
which was recently introduced into TiGL.
Compared the the former approach of manual implementation, automatic code generation has many benefits, e.g.
\begin{itemize}
\item reduced development time,
\item changes to CPACS can be adapted much faster,
\item fewer errors, and
\item CPACS files can now be checked strictly to their standard definition.
\end{itemize}
This generator was mainly developed by RISC Software GmbH
and can be publicly accessed on Github \cite{cpacs_generator}.
\textit{CPACSGen} is a command line tool, which uses the CPACS schema file
as its input and produces a C++ class for each of the CPACS nodes.
There are several ways to influence the code generation process.
For example, there are many node definitions in CPACS that are not used by TiGL.
Therefore, the generator allows the definition of a \textit{prune list} input file
that lists CPACS nodes, which are completely discarded -- including their sub-trees.
This is an effective mechanism to drastically reduce the amount of created code.
Manual modifications to the auto-generated code can be realized
by inheriting from the automatically generated classes, 
as the auto-generated code should not be modified by hand.
Although, CPACSGen was designed primarily for TiGL,
it can be used for other XML schema as well with only small adaptations.

\subsection{Software Bindings}
\label{sec:bindings}

Even though TiGL is written in C++,
it is mainly used by our users from the Python programming language.
In addition, TiGL also comes with bindings for C, MATLAB and Java.
Based on the public C interface of TiGL,
the bindings are automatically generated by a small self-developed tool included in TiGL.
The tool parses the C API and creates the bindings code for each of
the programming languages.
Sometimes, a C function signature can be ambiguous.
For example a pointer to an integer \texttt{int*} could be either an integer return
value or an input integer array. 
To overcome the ambiguity, annotations inside a function's docstring allow to define the semantics
of each function argument.
For each supported programming language, a code generator finally
produces the bindings code.
Right now, we are using the following technologies to enable the bindings:
\begin{itemize}
\item \textbf{Python:} Dynamic function calls via Python's ctypes.
\item \textbf{MATLAB:} One compiled MATLAB-mex file combined with a \texttt{*.m} file per function.
\item \textbf{JAVA:} Dynamic DLL loading with the JNA library \cite{jna}.
\end{itemize}

In addition to the relatively simplistic language bindings via
the public C API,
we started to bind the entire internal C++ interface to Python.
This makes it possible to use TiGL and pythonOCC \cite{pythonocc} --
the Python bindings to OpenCASCADE -- in an interoperable fashion.
This way, geometric objects can be passed from TiGL to OpenCASCADE and vice versa.
Just like pythonOCC, these bindings are created with SWIG \cite{swig}.
We experienced, that these new Python bindings encourage now also
external developers with no C++ knowledge to contribute code to TiGL.

\section{Geometry Module} \label{sect:geometry}
\subsection{B-spline modeling}
\begin{figure}[t!]
  \centering
  \includegraphics[height=5cm]{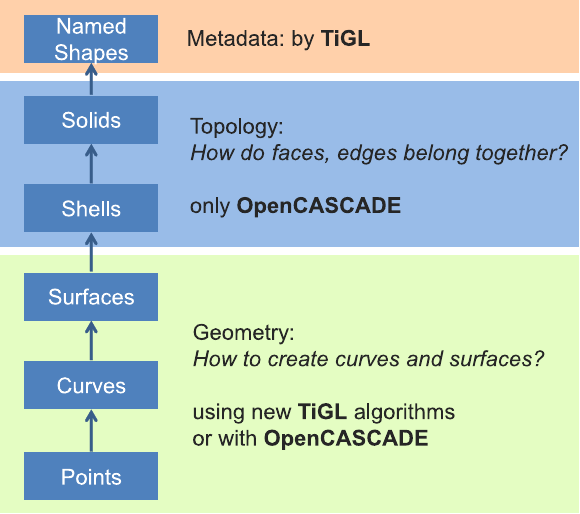}
  \caption{Bottom-up modeling of the geometries.}
  \label{fig:geometry_pipeline}
\end{figure}

TiGL uses the OpenCASCADE Technology CAD kernel \cite{opencascade} extensively for many tasks,
e.g. to create solid objects, apply Boolean operations to them,
and as a basis for the import and export modules.
The geometric operations in OpenCASCADE however are not used in
TiGL since several robustness issues were experienced in the past and the
quality of the generated surfaces was not always satisfying.
Therefore, most of the geometric modeling algorithms are implemented
in the TiGL software itself.
Just as OpenCASCADE, the geometric modeling is based on B-spline
curves and surfaces.
A B-spline curve is defined as
\begin{equation}
c(u) = \sum_{i=0}^{n-1} \vec{p}_i N_i^d(u, \boldsymbol {\tau}),
\end{equation}
where $\lbrace{\vec{p}_i\rbrace}$ are the $n$ control points of the curve,
$\boldsymbol{\tau}$ is its knot vector, 
and $\lbrace{N_i^d(u,\boldsymbol{\tau})\rbrace}$ are the B-spline basis function of degree $d$.
B-spline surfaces are defined as a tensor product of the B-spline basis functions,
in particular
\begin{equation}
s(u,v) = \sum_{i=0}^{n-1}\sum_{j=0}^{m-1} \vec{P}_{i,j} N_i^{\mu}(u, \boldsymbol {\tau_u}) N_j^{\nu}(v, \boldsymbol {\tau_v}).
\end{equation}
All geometric shapes in TiGL, such as wings, fuselages, or engines
have to be modeled as a combination of multiple B-spline surfaces.
A bottom-up approach, as proposed in \cite{haimes}, is used for the geometric modeling:
CPACS defines parametric points which are then used to 
build up curves, such as airfoils or fuselage sections.
These curves are then connected to create the final surfaces.
Several surfaces are eventually formed to solids and enriched
by meta-data, which contain additional information such as face names.
The single solid components are finally used to create the
shape of the entire aircraft via Boolean operations,
typically done in Constructive Solid Geometry.
This approach
is illustrated in \fref{fig:geometry_pipeline}.

The B-spline modeling algorithms used in TiGL are standard
methods from textbooks \cite{thenurbsbook, cagd}.
The most often used algorithm to create curves is B-spline interpolation,
which creates as B-spline curve that passes through a set of points.
Let $\lbrace{\vec{c}_i | i = 0 \dots n-1\rbrace}$ be a set of points
that should be interpolated at the curve parameters $\lbrace{u_i\rbrace}$,
then the interpolation conditions form the following linear system:
\begin{equation}
c(u_i) = \sum_{i=0}^{n-1} \vec{p}_i N_i^d(u_i, \boldsymbol {\tau}) = \vec{c}_i
\label{eq:curve_interpolation}
\end{equation}
This can be solved using standard linear solver methods, such as Gaussian elimination.
When interpolating a closed set of points -- i.e. where the first and last point
is identical -- 
with a periodic, $C^2$ continuous B-spline curve, this linear system can
get singular for even polynomial degrees.
To overcome this issue, the shifting method \cite{park} is used.
The interpolation of curves -- often referred to as surface skinning \cite{skinning1, skinning2} --
is similar to the curve interpolation.
Surface skinning requires a set of compatible B-spline curves,
where the curves differ only in their control points.
The control points of the skinning surface are then computed
by interpolating the curve's control points as before in equation~\eqref{eq:curve_interpolation}.

In addition to curve and surface interpolation,
TiGL also uses B-spline approximation algorithms in a few cases.
These algorithms perform a least-squares fit of a B-spline to a set of points.

\subsection{Curve Network Interpolation}
\label{sect:curve_network_interpolation}

At the heart of the geometric module is the curve network interpolation algorithm.
It allows an accurate modeling of surfaces while keeping the number of input curves small.
Compared to the simpler surface skinning method \cite{skinning1, skinning2},
where a set of profile curves is interpolated by a surface,
additional guide curves -- sometimes also called rail curves -- provide more control
over how the surface interpolates the profiles.
The algorithm is based on the Gordon surface method (see \Sref{sect:gordon_surfaces}),
which a almost never found in free or open source software.
To our knowledge, only \emph{Ayam} \cite{ayam} and Sintef's \emph{Go Tools} \cite{gotools}
implement the method,
but without addressing the curve network compatibility problem (see \Sref{sec:compatibility}).

Consider a curve network with $N$ profile curves $f_i(u): \mathbb R \rightarrow \mathbb R^3$ with $i=1\dots N, u \in [0,1]$ and
$M$ guide curves $g_j(v): \mathbb R \rightarrow \mathbb R^3$ with $j=1\dots M, v \in [0,1]$.
The curve network should be properly closed by surrounding profile and guide curves, i.e.

\begin{align}
\exists v_i, v'_i \in [0,1]&: f_i(0) = g_1(v_i) \wedge f_i(1) = g_M(v'_i) \quad &\forall i = 1 \dots N \\
\exists u_j, u'_j \in [0,1]&: g_j(0) = f_1(u_j) \wedge g_j(1) = f_N(u'_j) \quad &\forall j = 1 \dots M.
\end{align}

\begin{figure}[tb]
	\centering
\begin{tikzpicture}[scale=1.0]
\node[inner sep=0pt] (russell) at (0,0)
 {
   \includegraphics[width=0.64\textwidth]{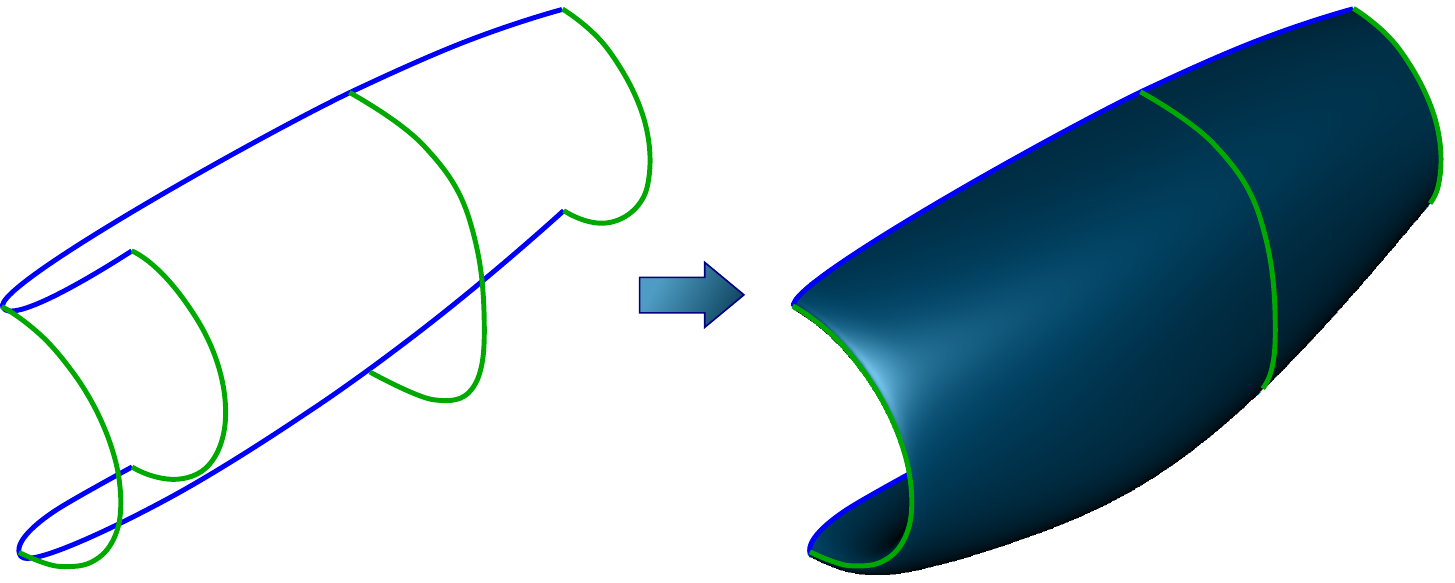}
 };
 \draw (-3.8, 1.3) node [text={rgb,255:red,0; green,0; blue,200}]{Profile $g(v)$};
 \draw (-2.7, 0.1) node [text={rgb,255:red,0; green,140; blue,0}]{Guide $f(u)$};
\end{tikzpicture}
\caption{Curve network interpolation via profile and guide curves.}
\label{fig:curve_network}
\end{figure}

This basically means, that all profile curves must begin at the first and end at the last guide curve.
And vice versa, all guide curves must begin at the first and end at the last profile curve.
Such a curve network is depicted in \fref{fig:curve_network}.
This curve network should now be interpolated by one single surface $s(u,v): \mathbb R \times \mathbb R \rightarrow \mathbb R^3$.
If it is enforced, that the input curves are iso-parametric curves of the resulting surface, 
the following conditions must hold:
\begin{align}
\exists v_i: s(u, v_i) &= f_i(u), \quad  i = 1 \dots N \\
\exists u_j: s(u_j, v) &= g_j(v), \quad  j = 1 \dots M
\end{align}
Using these iso-parametric conditions, it is now easy to see,
that for any profile curve $f_i$ and guide curve $g_j$,
the \emph{compatibility condition} of the curve network follows:
\begin{equation}
f_i(u_j) = s(u_j, v_i) = g_j(v_i),\quad	i = 1 \dots N,\;j = 1 \dots M \; ,
\label{eq:compat_conditions}
\end{equation}
i.e. all profile curves must intersect a guide curve at the same curve parameter,
and all guides curves must intersect a profile curve at the same parameter.

\subsubsection{Gordon Surfaces}
\label{sect:gordon_surfaces}
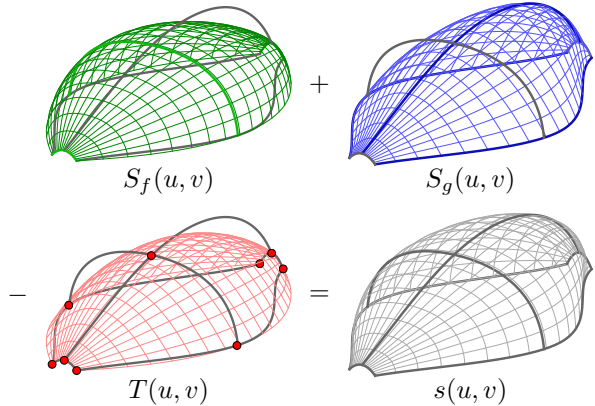
\begin{figure}[t!]
	\centering
\begin{tikzpicture}[scale=0.80]
\node[inner sep=0pt] (russell) at (0,3.5)
 {
   \includegraphics[width=0.21\textwidth, trim=0 0 0 0, clip]{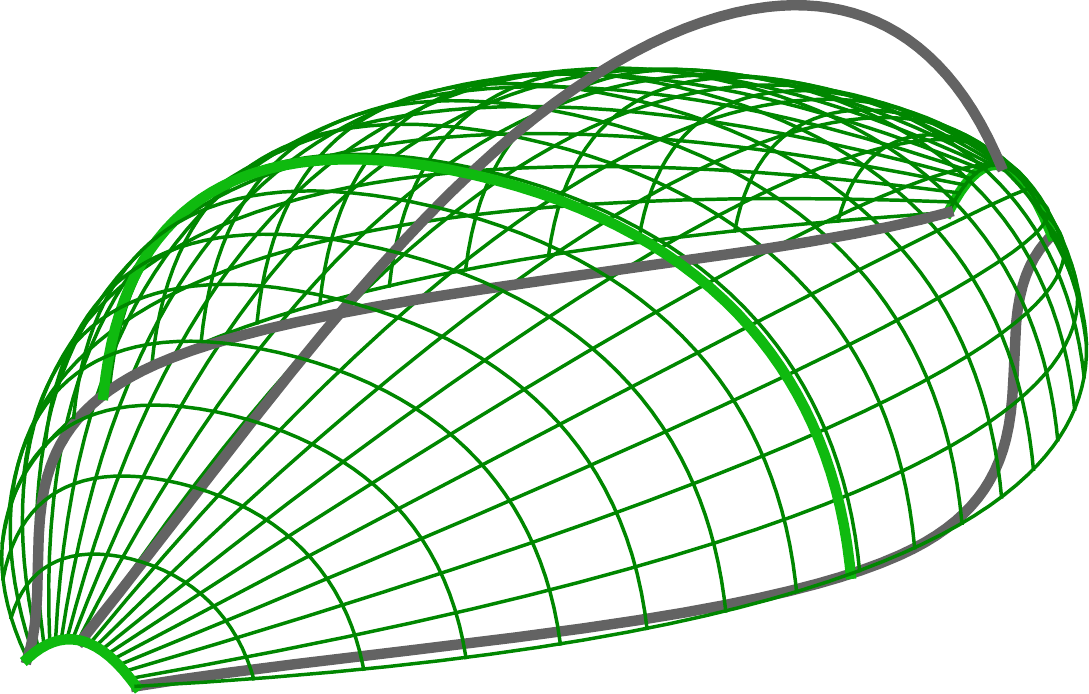}
 };
 \draw (0, 1.9) node {$S_f(u,v)$};
 \draw (2.5, 3.5) node {$+$};
\node[inner sep=0pt] (russell) at (5,3.5)
 {\includegraphics[width=0.21\textwidth, trim=0 0 0 0, clip]{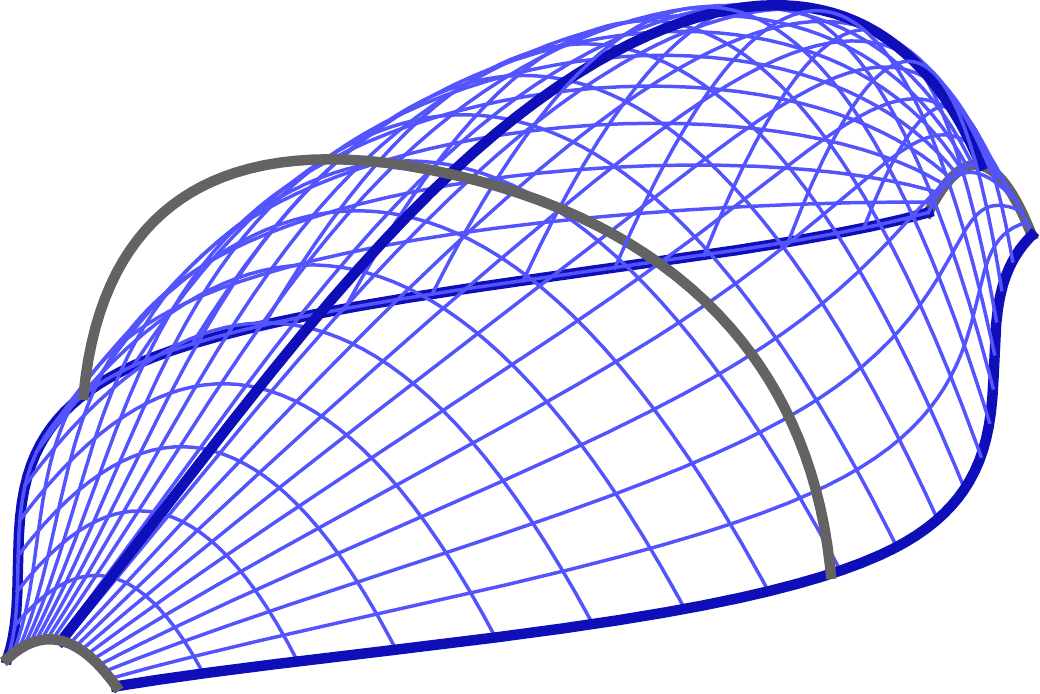}};
 \draw (5, 1.9) node {$S_g(u,v)$};
 \draw (-2.5, 0) node {$-$};
\node[inner sep=0pt] (russell) at (0,0)
 {\includegraphics[width=0.21\textwidth, trim=0 0 0 0, clip]{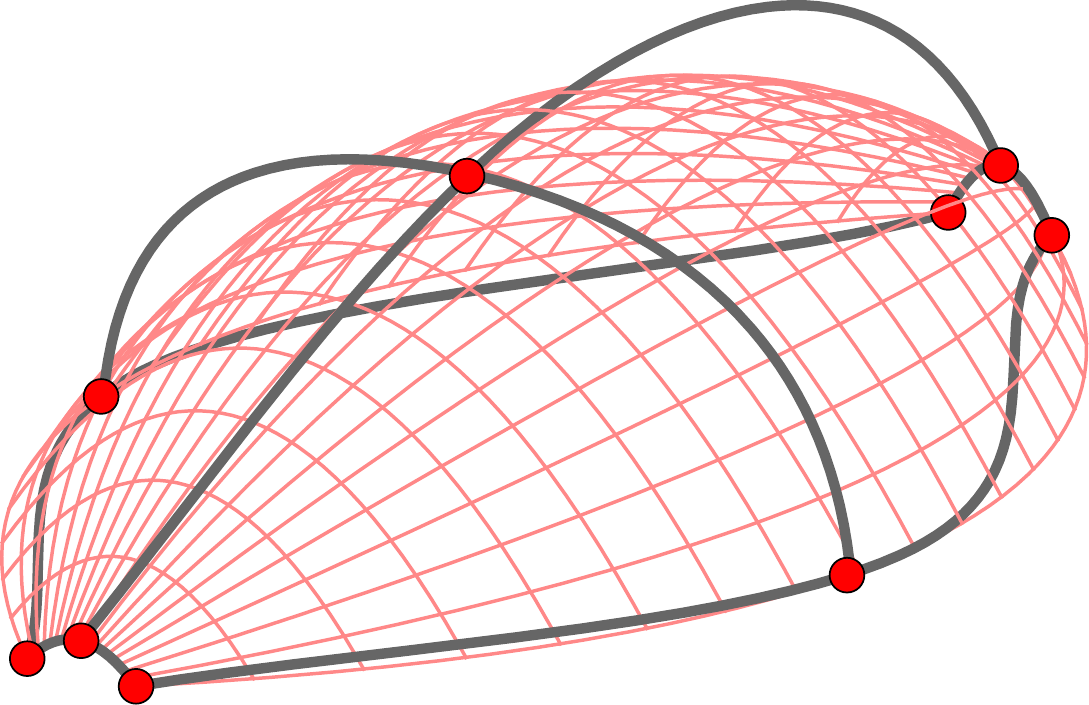}};
  \draw (0, -1.6) node {$T(u,v)$};
\draw (2.5, 0) node {$=$};
\node[inner sep=0pt] (russell) at (5,0)
 {\includegraphics[width=0.21\textwidth, trim=0 0 0 0, clip]{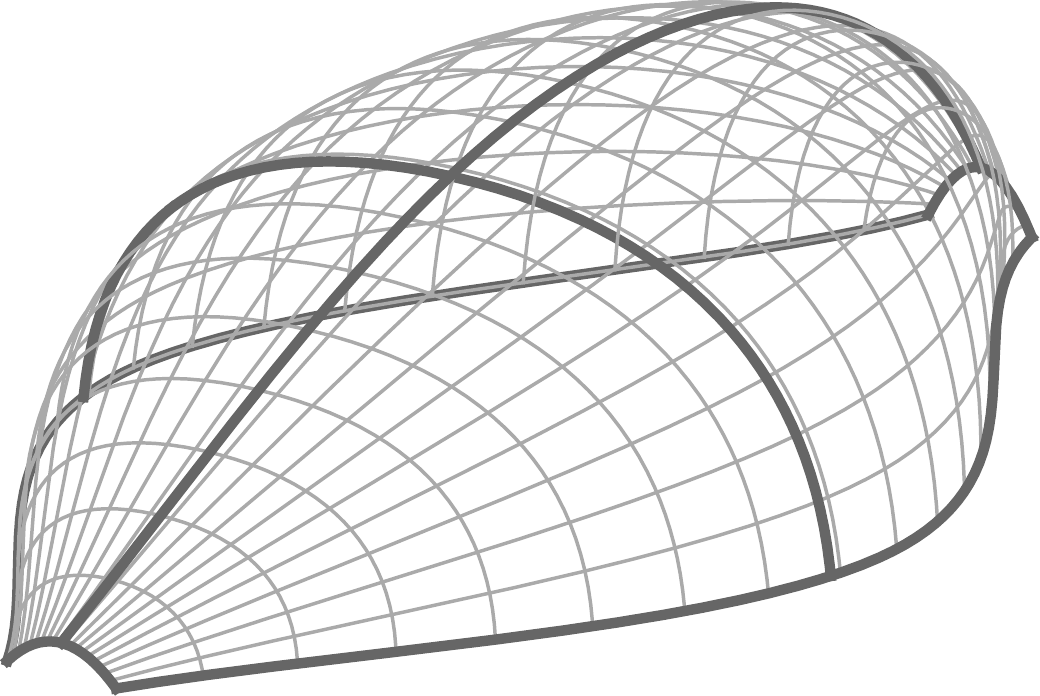}};
 \draw (5, -1.6) node {$s(u,v)$};

\end{tikzpicture}
\caption{Construction of the Gordon surface via its three interpolation surfaces.}
\label{fig:gordon_construction}
\end{figure}

 William J. Gordon published a method \cite{gordon1969spline} that is able to interpolate a curve network
 if it fulfills the curve compatibility condition \eqref{eq:compat_conditions}:
 For any set of spline blending functions $\lbrace{\phi_j(u)\rbrace}$ and $\lbrace{\psi_i(v)\rbrace}$
 satisfying the conditions
 
\begin{equation}
\phi_j(u_k) = 
  \begin{cases} 
      0 & k \neq j \\
      1 & k = j 
  \end{cases}  \text{ and }
\psi_i(v_k) = 
  \begin{cases} 
      0 & k \neq i \\
      1 & k = i 
  \end{cases},
\end{equation}
the following blending surface interpolates the curve network $\lbrace{ f_i(u)\rbrace}$ and $\lbrace{ g_j(v)\rbrace}$:

\begin{equation}
s(u,v) = \sum^N_{i=1} f_i(u) \psi_i(v) + \sum^M_{j=1} g_j(v) \phi_j(u) - \sum^N_{i=1}\sum^M_{j=1} \vec  \alpha_{i,j} \phi_j(u)\psi_i(v)
\label{eq:gordon}
\end{equation}
Here $\vec  \alpha_{i,j}$ is the intersection point of the i-th profile curve $f_i(u)$ with the j-th guide curve $g_j(v)$,
i.e. 
\begin{equation}
\vec  \alpha_{i,j} = f_i(u_j) = g_j(v_i).
\end{equation}
Equation \eqref{eq:gordon} can now be rewritten as follows:

\begin{equation}
s(u,v) = S_f(u,v) + S_g(u,v) - T(u,v)
\end{equation}
Each of the three summands can be interpreted as an interpolation surface.
The first $S_f(u,v)$ is a surface that interpolates the profile curves $\lbrace{f_i(u)\rbrace}$,
whereas the second term $S_g(u,v)$ is an interpolation surface for the guide curves $\lbrace{g_j(v)\rbrace}$.
The third term, often also called tensor product surface, interpolates the net of intersection points  $\lbrace{\vec \alpha_{i,j}\rbrace}$.
\fref{fig:gordon_construction} illustrates the surface construction principle for Gordon surfaces. 
It can be seen as a generalization of the Coons-patch method \cite{coons1967surface} 
to more than two profile and guide curves.

\subsubsection{Gordon Surfaces with B-splines}
Since TiGL relies on the B-spline based OpenCASCADE CAD library,
all curves of the curve network are B-splines and the Gordon surface
must also be a B-spline surface finally. 
Since the Gordon surface consists of two skinning surfaces and one tensor product surface,
the surfaces $S_f(u,v)$ and  $S_g(u,v)$ can be interpreted as the B-spline
based skinning surfaces of the profile curves and guide curves.
According to ``the NURBS Book'' \cite[p.~485]{thenurbsbook},
the blending functions $\lbrace{\phi_j(u)\rbrace}$ and $\lbrace{\psi_i(v)\rbrace}$ can be interpreted as B-spline basis functions.

For the further derivation of the B-spline based Gordon surface method, it is required
that all profile curves share
\begin{itemize}
\item the same degree
\item and a common knot vector
\end{itemize}
in addition to the compatibility conditions of the curve network \eqref{eq:compat_conditions}.
The same should also apply to all guide curves.
In practice, both can always be achieved by using degree elevation \cite{degree_elevation1, degree_elevation2}
and knot insertion \cite{knot_insert_boehm, knot_insert_oslo} of the input curves.

In this case, all profile curves $\lbrace{f_k(u)\rbrace}$ and all guide curves $\lbrace{g_l(v)\rbrace}$ are of the form

\begin{align}
f_k(u) &= \sum_{i=0}^{n-1} \vec{p}_i^{(k)} N_i^\nu(u, \boldsymbol {\tau_f}),\quad k=1 \dots N  \nonumber \\
g_l(v) &= \sum_{i=0}^{m-1} \vec{q}_i^{(l)} N_i^\mu(v, \boldsymbol {\tau_g}),\quad l=1 \dots M.
\end{align}
Here,  $\lbrace{\vec{p}_i^{(k)}}\rbrace$ are the control points of the k-th profile curve
and $\lbrace{\vec{q}_i^{(l)}\rbrace}$ are the control points of the l-th guide curve.

If the profile curves $\lbrace{f_k(u)\rbrace}$ are skinned with a B-spline surface with knot vector $\boldsymbol{\xi_f}$ and degree $d_f$
and the guide curves $\lbrace{g_l(v)\rbrace}$ are skinned with a B-spline surface with knot vector $\boldsymbol{\xi_g}$ and degree $d_g$,
Gordon's equation \eqref{eq:gordon} for B-splines can be rewritten as follows:
\begin{align}
s(u,v) =& \sum_{i=0}^{n-1} \sum_{j=0}^{N-1} \vec P_{i,j} N_i^\nu(u, \boldsymbol {\tau_f})  N_j^{d_f}(v, \boldsymbol {\xi_f}) \nonumber \\
       &+ \sum_{j=0}^{m-1} \sum_{i=0}^{M-1} \vec Q_{i,j} N_i^{d_g}(u, \boldsymbol {\xi_g}) N_j^\mu(v, \boldsymbol {\tau_g}) \nonumber \\
       &- \sum_{i=0}^{N-1} \sum_{j=0}^{M-1} \vec T_{i,j} N_i^{d_g}(u, \boldsymbol {\xi_g}) N_j^{d_f}(v, \boldsymbol {\xi_f})
\label{eq:gordon_bspline}
\end{align}
The three control nets $\lbrace{\vec P_{i,j}\rbrace}$, $\lbrace{\vec Q_{i,j}\rbrace}$ and $\lbrace{\vec T_{i,j}\rbrace}$ must comply
with the following interpolation conditions for all $k = 1 \dots N$ and $l = 1 \dots M$:
\begin{align}
\sum_{j=0}^{N-1} \vec P_{i,j} N_j^{d_f}(v_k, \boldsymbol{\xi_f}) &= \vec p_i^{(k)}, \quad i = 0 \dots n-1 \label{eq:interp_profiles}\\
\sum_{i=0}^{M-1} \vec Q_{i,j} N_i^{d_g}(u_l, \boldsymbol{\xi_g}) &= \vec q_j^{(l)}, \quad j = 0 \dots m-1 \label{eq:interp_guides}\\
\sum_{i=0}^{N-1} \sum_{j=0}^{M-1} \vec T_{i,j} N_i^{d_g}(u_l, \boldsymbol {\xi_g}) N_j^{d_f}(v_k, \boldsymbol {\xi_f}) &= \vec \alpha_{l,k} \label{eq:interp_intersection}
\end{align}
These ensure, that the first term interpolates the profile curves, the second term interpolates the guides curves,
and the third term interpolates the curve network's intersection points  $\lbrace{\vec \alpha_{l,k}\rbrace}$.
The interpolation conditions are linear systems, which can be solved again using e.g. Gaussian elimination.
It should be emphasized that the interpolation of the intersection points \eqref{eq:interp_intersection} must
use the same interpolation parameters, degrees and knot vectors as the interpolation of the profile curves \eqref{eq:interp_profiles}
and the guide curves \eqref{eq:interp_guides}.

The B-spline based Gordon surface \eqref{eq:gordon_bspline} still is a superposition of the three interpolation surfaces.
The three surfaces differ in their degree and knot vector.
To be usable for TiGL in the end, the Gordon surface has to be converted back to a single B-spline surface.
Fortunately, this is easy to achieve:
first the degree of the surfaces is elevated to their maximum u- and v-degree.
Then, knots in u- and v-direction have to be inserted, such that all surfaces
share the same knot vector.
After degree elevation and knot insertion, the three surfaces are compatible
and thus have the same number of control points.
The final B-spline based Gordon surface is created by adding the control points of the skinning surfaces
and subtracting the control points of the tensor product surface.

\subsubsection{Achieving compatibility of the curve network}
\label{sec:compatibility}
Until know, it was assumed that the curves of the curve network are compatible,
i.e. they meet the compatibility conditions \eqref{eq:compat_conditions}.
In practice, however, this is almost never the case, since the curves can be parametrized arbitrarily.
To meet the compatibility conditions, the curve network has to be reparametrized first.

Without loss of generality, the following derivations will be performed
on the profile curves.
For the guide curves, the equations then follow analogously.
Let the original profile curve $\tilde f_k(\tilde u)$ intersect the
original guide curve $\tilde g_l(\tilde v)$ at parameter $\tilde u_{l,k}$, i.e.
\begin{equation}
\tilde f_k(\tilde u_{l,k}) = \tilde g_l(\tilde v_{k,l}).
\label{eq:intersection}
\end{equation}
It is known from the analyses before, that the profile curve has to intersect
all guide curves at the parameters $\lbrace{u_l\rbrace}$.
Using a reparametrization function $\sigma_k(u)$, the reparametrized profile curve $f_k(u)$ can now be defined as
\begin{equation}
f_k(u) = \tilde f_k(\sigma_k(u)).
\end{equation}
The function $\sigma_k(u)$ must satisfy
\begin{equation}
\sigma_k(u_l) = \tilde u_{l,k}.
\label{eq:repara_mapping}
\end{equation}
This type of B-spline reparametrization is described in \cite[pp. 241]{thenurbsbook} as ``internal point mapping".
The choice of the reparametrization function $\sigma_k(u)$ is arbitrary
but must fulfill in addition to \eqref{eq:repara_mapping} the following conditions:
\begin{enumerate}
\item It must be twice continuously differentiable since curvature continuous surfaces are required in the end.
\item The mapping function must be strictly increasing ($\sigma_k^\prime(u) > 0$)
      such that each original parameter $\tilde u$ maps
      uniquely to one target parameter $u$ (monotone + bijective).
\end{enumerate}
For the sake of simplicity, a B-spline interpolation based reparametrization
of degree 3 was chosen, which is
\begin{equation}
\tilde u = \sigma_k(u) = \sum_i \vec \Psi_{i,k} N_i(u, \boldsymbol \tau_\sigma),\quad\mathrm{such \, that}\quad
\tilde u_{l,k} = \sum_i \vec \Psi_{i,k} N_i(u_l, \boldsymbol \tau_\sigma), 
\end{equation}
where the second part ensures the mapping \eqref{eq:repara_mapping}.
Here, $\vec \Psi_{i,k} \in \mathbb R$ are the control values of the interpolation
B-spline of the curve $\sigma_k(u)$.
It should be noted, that if original parameters  $\lbrace{\tilde u_{l,k}\rbrace}$ and
target parameters $\lbrace{u_l\rbrace}$ deviate too much,
the function $\sigma_k(u)$ can lose its bijectivity due to B-spline oscillations.
For now, the bijectivity is checked by the TiGL software.
If it fails, the code will create an error.

After the reparametrization function $\sigma_k(u)$ has been found,
the composed curve $\tilde f_k(\sigma_k(u))$ has to be 
transformed to B-spline form.
There are two ways to achieve this:
\begin{enumerate}
\item Exact reparametrization as described in \cite[pp. 247]{thenurbsbook}.
      Since it is exact, it does not introduce any error to the method.
      The big drawback of this method is the greatly increased
      degree of the profile / guide curve after reparametrization. 
      Since the degrees of the original curve and the reparametrization function
      are multiplied for the resulting curve,
      an input curve $\tilde f(\tilde u)$ with degree
      4 and a reparametrization function $\sigma(u)$ of degree
      3 results in a profile function of degree 12.
      As a consequence, also the degree of the final surface will be large.
\item Approximate reparametrization: The reparametrized function is approximated
      by sampling the curve $\tilde f_k(\sigma_k(u))$ and subsequently creating a new
      one from these sampled points.
      This way, the degree can be limited and the knot vector can
      be chosen more freely.
      The obvious drawback is the additional error introduced by the approximation.
\end{enumerate}
For TiGL, the second method was chosen due to the following reasons:
First, large degrees of the final surfaces should be avoided to
keep the numerical complexity low.
Second, the error of the approximation can be controlled and
it can always be reduced by increasing the number of control points.
Third, the free choice of the knot vector can be exploited such that
e.g. the resulting profile curves $\lbrace{f_k(u)\rbrace}$ all have the same knot vector.
This helps to keep the number of knots and control points of the final surface low. 

When using the approximation technique, it is essential that the
re\-para\-me\-trized curve still exactly passes through its intersection
points of the curve network,
such that Gordon's equation \eqref{eq:gordon} is still valid.
To achieve this, a hybrid approximation / interpolation technique
of the sampled curve points is used: Let $\lbrace{\hat u_j\rbrace}$
be $n_s$ sample parameters of the curve $\tilde f_k(\sigma_k(u))$ and
let $\lbrace{u_l\rbrace}$ be the curve intersection parameters
of the curve with the curve network.
Then the control points $\lbrace{p^{(k)}_i\rbrace}$ of the approximation B-spline are computed
by solving the following constrained linear least squares problem:
\begin{align}
\min_{p^{(k)}} \sum_{j=0}^{n_s-1} \left[\sum_{i=0}^{n-1} p^{(k)}_i N^\nu_i(\hat u_j, \boldsymbol {\tau_f}) - \tilde f_k\left(\sigma_k(\hat u_j)\right)\right]^2,  \nonumber \\
s.t.\quad \sum_{i=0}^{n-1} p^{(k)}_i N^\nu_i(u_l, \boldsymbol{ \tau_f}) = \tilde f_k(\sigma_k(u_l)), \quad l = 1 \dots M
\label{eq:approx_interp_min}
\end{align}
Using the Lagrange multiplier method, this problem \eqref{eq:approx_interp_min}
can be transformed into the constrained normal equation:

\begin{equation}
\begin{bmatrix}
  \boldsymbol{\hat{N}}^T\boldsymbol{\hat{N}} & \boldsymbol{N}^T  \\
  \boldsymbol{N}& \boldsymbol{0} 
\end{bmatrix} 
 \cdot
\begin{bmatrix}
  \boldsymbol{p}  \\
  \boldsymbol{\lambda} 
\end{bmatrix} 
 = 
\begin{bmatrix}
  \boldsymbol{\hat{N}}^T \boldsymbol{\hat{c}}  \\
  \boldsymbol{c} 
\end{bmatrix},
\label{eq:approx_interp}
\end{equation}
with
\begin{align*}
\hat{N}_{ji} &:= N_i^\nu(\hat u_j, \boldsymbol {\tau_f}), \quad
   \hat{c}_j :=  \tilde f_k(\sigma_k(\hat u_j)), \\
         N_{li} &:= N_i^\nu(u_l, \boldsymbol {\tau_f}), \quad 
         c_l :=  \tilde f_k(\sigma_k(u_l)), \\
         \text{and } &i = 0 \dots n-1, \; j = 0 \dots n_s - 1, \; l = 1 \dots M. 
\end{align*}
As usual, $\boldsymbol{\lambda} $ represents the Lagrange multipliers of the constraint problem.
The linear system \eqref{eq:approx_interp} is finally solved using Gaussian elimination.
If the original curve contains kinks, these kinks are reproduced
in the reparametrized curve by inserting knots with a multiplicity
of the curve's degree $\nu$ into  the knot vector $\boldsymbol {\tau_f}$
prior to the approximation.

The approximation can be performed with an arbitrarily chosen number of sample points $n_s$.
To get a unique solution, $n_s$ must be larger than the number
of control points $n$ of the reparametrized curve.
The number of control points should be as small as possible to avoid
unnecessary computational complexity but should be large enough
to keep the approximation error small.
In TiGL, we simply use roughly the same number of control points $n$ for the
reparametrized curve as for the original curve.
This way, it is possible to reproduce all features of the original curve.
The knot vector $\boldsymbol {\tau_f}$ is chosen to be uniform.

\subsubsection{Algorithm}

\begin{algorithm}[b]
\caption{B-spline based Gordon surface creation algorithm.}
\label{algo:gordon_bspline}
\DontPrintSemicolon
\KwIn{Curve network of profiles $\tilde{f}_k(\tilde u)$ and guides $\tilde{g}_l(\tilde v)$}
\KwOut{The Gordon surface $s(u,v)$ that interpolates the network}
Compute intersections parameters  $\tilde u_{l,k}$ and $\tilde v_{k,l}$ such that \eqref{eq:intersection} holds.\;
\For{$l = 1 \dots M$}{
	$u_l \gets \frac{1}{N}\sum_k \tilde u_{l,k}$\;
}
\For{$k = 1 \dots N$}{
	$v_k \gets \frac{1}{M}\sum_l \tilde v_{k,l}$\;
}
$n \gets $ maximum number of control points of the profile curves $\tilde{f}_k(\tilde u)$\;
$m \gets $ maximum number of control points of the guide curves $\tilde{g}_l(\tilde v)$\;
\For{$k = 1 \dots N$}{
	Compute $f_k(u)$ by reparametrizing $\tilde{f}_k(\tilde u)$ using $n$ control points
	and original intersection parameters $\lbrace{\tilde u_{l,k}\rbrace}$ and target parameters $\lbrace{u_l\rbrace}$ as described in \Sref{sec:compatibility}.\;
}
\For{$l = 1 \dots M$}{
	Compute $g_l(v)$ by reparametrizing $\tilde{g}_l(\tilde v)$ using $m$ control points analogous to the profiles.\;
}
Compute profile skinning surface $S_f(u,v)$ with interpolation parameters
  $\lbrace{v_k\rbrace}$ according to \eqref{eq:interp_profiles}.\;
Compute guide skinning surface $S_g(u,v)$ with interpolation parameters
  $\lbrace{u_l\rbrace}$ according to \eqref{eq:interp_guides}.\;
Compute tensor product surface $T(u,v)$ with interpolation parameters
  $\lbrace{u_l\rbrace}$ and $\lbrace{v_k\rbrace}$ according to \eqref{eq:interp_intersection}.\;
Make $S_f(u,v)$, $S_g(u,v)$ and $T(u,v)$ compatible by degree elevation and knot insertion.\;
Create final surface $s(u,v)$ by adding/subtracting the control points of the compatible interpolation surfaces.\;
\Return{$s(u,v)$}\;
\end{algorithm}

\begin{figure}[t]
\centering
\subcaptionbox{Wing: 4 profiles, 3 guides, with kink}
{
  \includegraphics[width=0.37\textwidth]{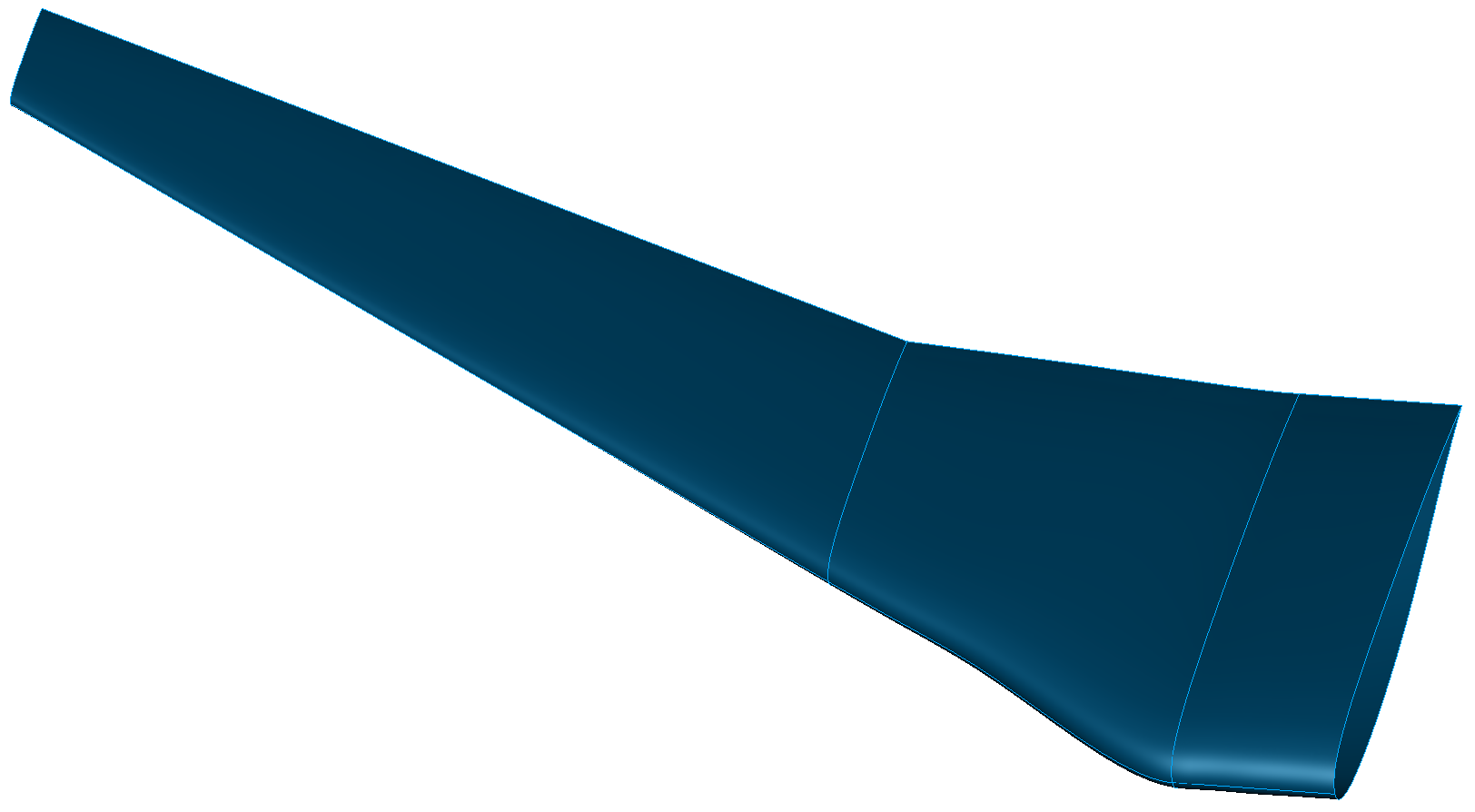}
}\qquad
\subcaptionbox{Spiral-shaped wing: 6 profile, 3 guides	}
{
  \includegraphics[width=0.37\textwidth]{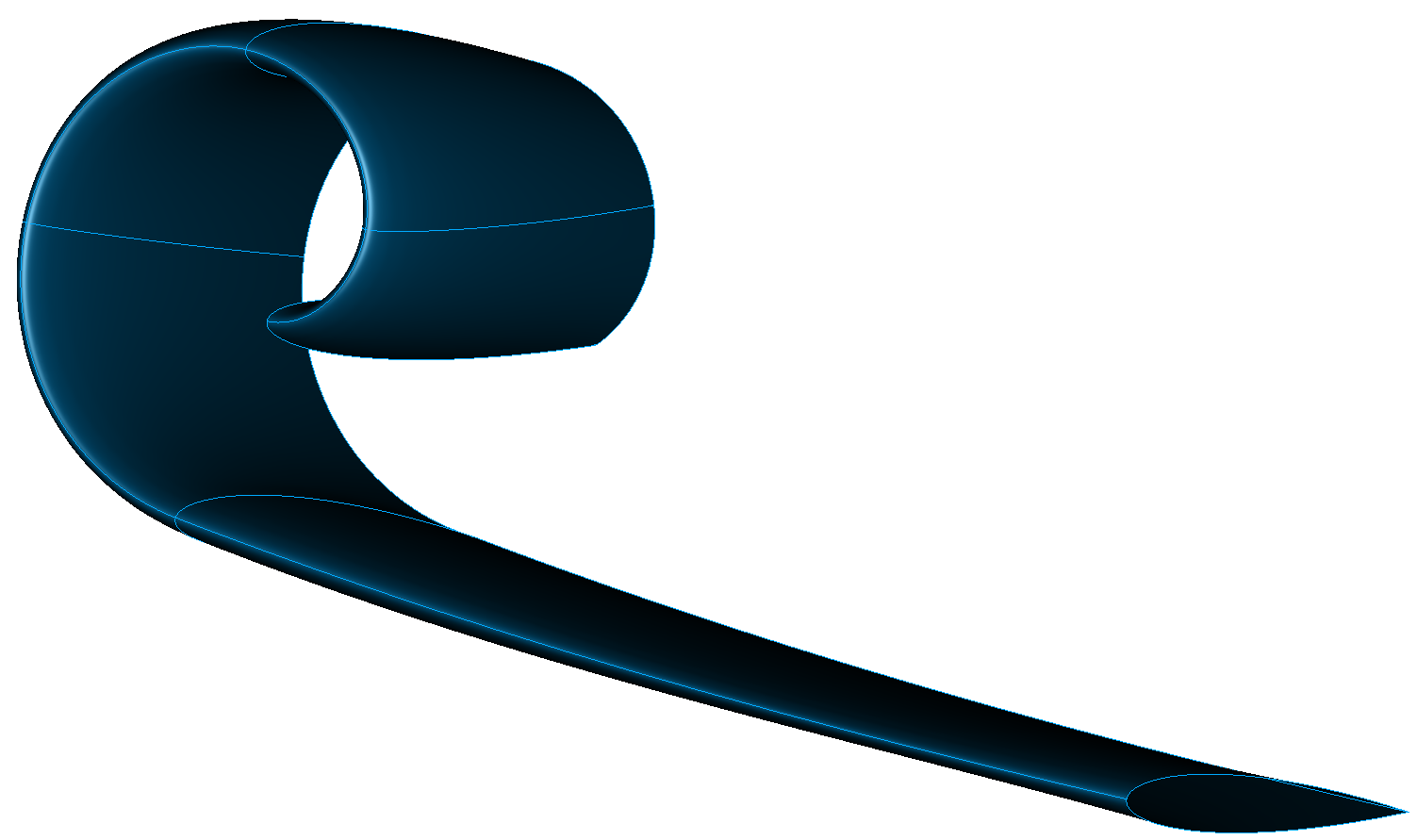}
}
\subcaptionbox{DLR-D150 fuselage: 10 profiles, 10 guides}
{
  \includegraphics[width=0.37\textwidth]{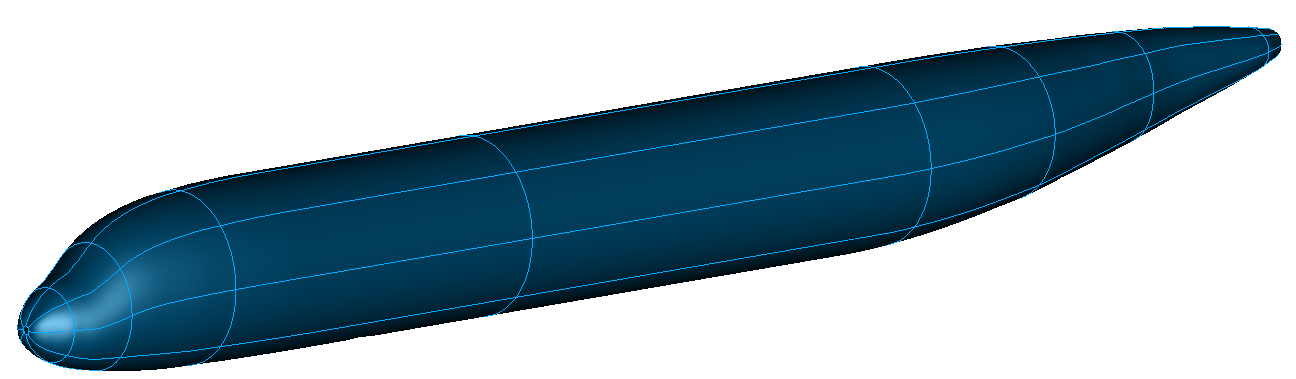}
}\qquad
\subcaptionbox{Belly fairing: 6 profiles, 2 guides}
{
  \includegraphics[width=0.37\textwidth]{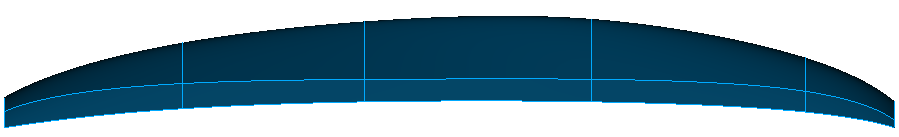}
}
\subcaptionbox{Extreme case helicopter: 6 profiles, 68 guide curves!\label{fig:gordon_heli}}
{
  \includegraphics[width=0.37\textwidth]{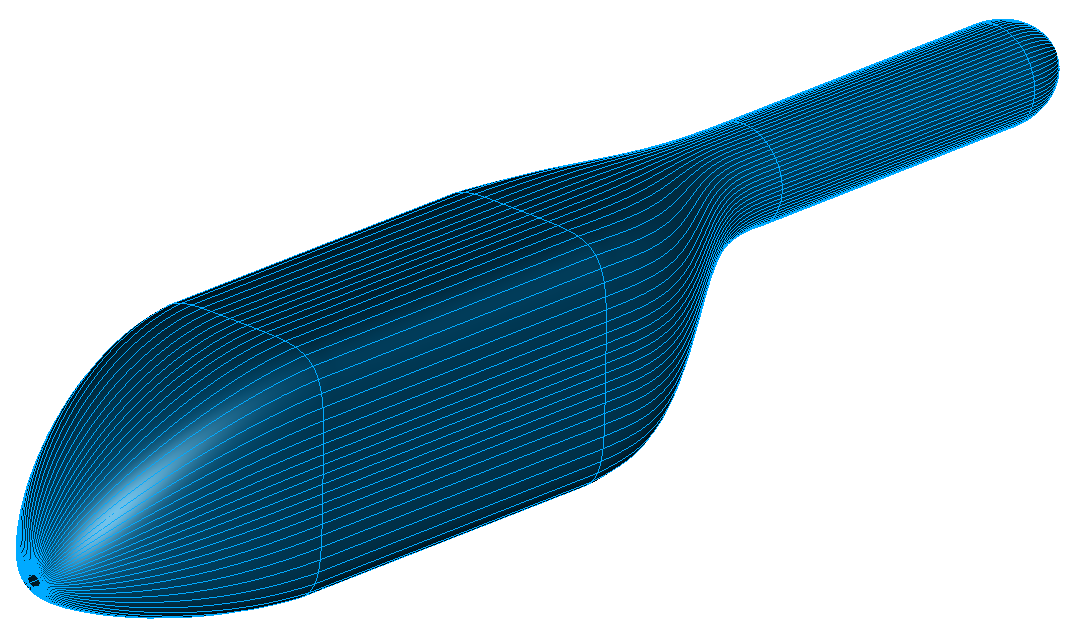}
}\qquad
\subcaptionbox{Engine nacelle: 5 profiles, 4 guides}
{
  \includegraphics[width=0.37\textwidth, trim=-150 -0 -150 -40, clip]{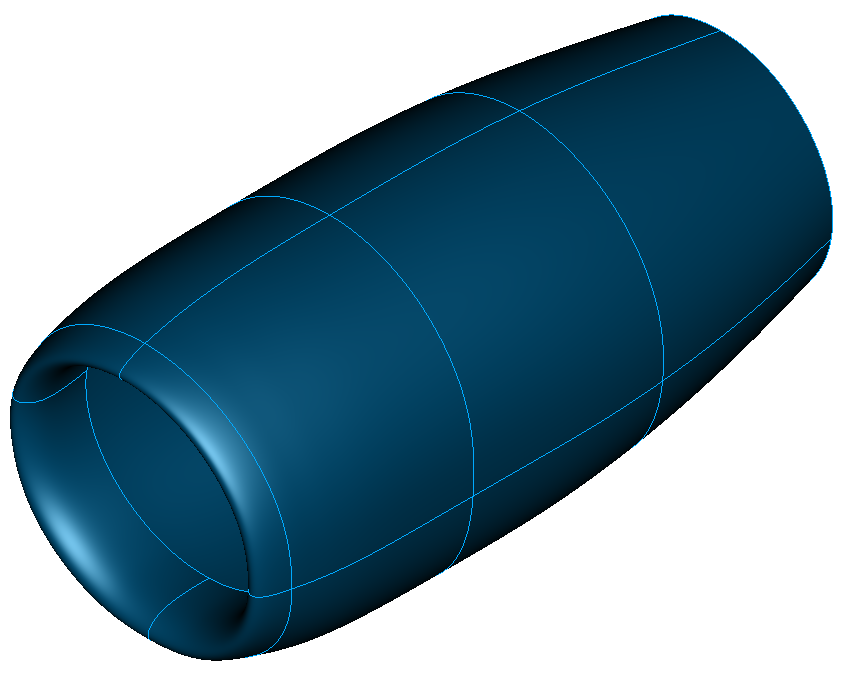}
}
\caption{Generated aircraft surfaces with the Gordon surface method. The helicopter
case demonstrates, that the method also works for a large curve network.}
\label{fig:gordon_examples}
\end{figure}

The whole algorithm combines the previously described steps.
To achieve a low number of control points of the final
surface, we use the same number of control points $n$ in all profile / guide curves
in the reparametrization step.
By using always the same number of control points,
all profiles / guide curves will get the same uniform knot vector
and hence also the skinning surface.
If the knot vectors were different, all knot vectors
would have to be merged first. 
This would result in a very large knot vector
and therefore a large number of control points for the final surface.
The pseudo code of our B-spline based
Gordon method is depicted in Algorithm~\ref{algo:gordon_bspline}.
\Fref{fig:gordon_examples} shows six different example geometries
that are created with this algorithm.
The extreme helicopter case (see \fref{fig:gordon_heli}) shows
that the algorithm is also suitable for very large curve networks.
The resulting surfaces are smooth -- except for intentionally inserted kinks --
and interpolate the curves as they should.

\section{Aircraft Component Module} \label{sect:aircraftComponentModeling}
Many aircraft component geometries can be generated using a network of profile and guide curves. 
\Sref{sect:profilesguides} will therefore describe how these curves can be defined according to the CPACS definition. 
Afterwards, details about the definition and modeling of wings, fuselages, control surfaces, structural elements as well as engine nacelles and pylons are presented.

\subsection{Profile and Guide Curves} \label{sect:profilesguides}
In this section, we are going to introduce the two basic building blocks for modeling wings and fuselages: profile and guide curves.
Both of these entities are defined with respect to some local coordinate system as it is common to the CPACS description.
For the wing profiles, TiGL implements a definition based on local points as well as a parametric description (CST).
The guide curves are defined by a set of points in a local coordinate system.

\subsubsection{Profiles from Point Lists}
\label{ssub:profiles_from_point_lists}
This is the most commonly used approach to creating wing and fuselage profiles with TiGL.
Given a set of three-dimensional points in a local coordinate system of a section, a B-spline curve is interpolated.
In a second step the curve is transformed to a global coordinate system to bring it to the position of the section and scale it accordingly.

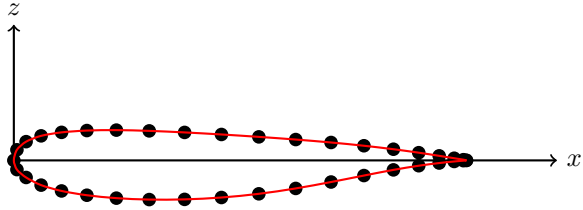
\begin{figure}[h]
    \centering
    \begin{tikzpicture}[scale=6]

\edef\points{}
\foreach \point [count=\i] in {
(1.0, 0.0),
 (0.99318065, 0.00156842),
 (0.97290862, 0.00478703),
 (0.93973687, 0.01009993),
 (0.89457025, 0.01704458),
 (0.83864078, 0.02471343),
 (0.77347407, 0.03240848),
 (0.7008477, 0.03960558),
 (0.62274273, 0.04610725),
 (0.54128966, 0.0520147),
 (0.45871032, 0.05732852),
 (0.37725724, 0.06187089),
 (0.29915227, 0.06527812),
 (0.22652591, 0.06697582),
 (0.1613592, 0.06600959),
 (0.10542973, 0.06200228),
 (0.06026312, 0.05427176),
 (0.02709137, 0.04144676),
 (0.00681934, 0.02344718),
 (0.0, 0.0),
 (0.00681934, -0.02033238),
 (0.02709137, -0.0377444),
 (0.06026312, -0.05467523),
 (0.10542973, -0.06753082),
 (0.1613592, -0.07701615),
 (0.22652591, -0.08345515),
 (0.29915227, -0.08659188),
 (0.37725724, -0.08618776),
 (0.45871032, -0.08202661),
 (0.54128966, -0.07386881),
 (0.62274273, -0.06198569),
 (0.7008477, -0.0477972),
 (0.77347407, -0.03348007),
 (0.83864078, -0.02095115),
 (0.89457025, -0.01177135),
 (0.93973687, -0.00582898),
 (0.97290862, -0.00238273),
 (0.99318065, -0.00057906),
 (1.0, 0.0)
} {
    \def\this{point-\i}
    \node[coordinate] (\this) at \point {} ;
    \fill (\this) circle (0.015) ;
    \xdef\points{(\this) \points}
}

\draw [<-,thick] (0.0,0.3) node (zaxis) [above] {$z$} -- (0.0,0.0) node (origin)  {};
\draw [->,thick] (0.0,0.0) node (origin) {} -- (1.2,0.0) node  (xaxis) [right] {$x$};
\draw [red, thick] plot [smooth] coordinates {\points};

\end{tikzpicture}
    \caption{Example for a wing profile created from a point list of $x$-$z$ coordinates.}
    \label{fig:profile_from_point_list}
\end{figure}

A typical airfoil can be seen in \fref{fig:profile_from_point_list}.
It is created by a list of $x$-$z$-points which are ordered in a mathematically positive sense. 
The list starts and ends at the trailing edge of the airfoil.

\subsubsection{Profiles from Parametrized Curves (CST)}
\label{ssub:profiles_from_parametrized_curves}
An alternative to the creation of wing profiles via points is the analytic description of the airfoils by the Class Shape Transformation method (CST)~\cite{cst}.
Both the upper and the lower half of the profile is defined by a two-dimensional curve, which reads
\begin{equation*}
    \zeta (\psi) = C^{N_1}_{N_2}(\psi) S_\mathbf{A}(\psi) + \psi \zeta_T, \; \quad \psi \in [0, 1].
\end{equation*}
The exponents $N_1$ and $N_2$ of the \emph{Class function}
$
    C^{N_1}_{N_2}(\psi) = \psi^{N_1} (1 - \psi)^{N_2} \; ,
$
determine the slope at the leading and trailing edge, respectively.
The \emph{Shape function} 
\begin{equation*}
    S_\mathbf{A}(\psi) = \sum_{i=0}^{n} A_i B_{i, n} (\psi)
\end{equation*}
is a linear combination of Bernstein basis functions 
$
B_{i, n}(\psi) = \binom{n}{i} \psi^i (1 - \psi)^{n - i}
$
of degree $n$ and controls the shape of the airfoil.
The size of the trailing edge is given by $\zeta_T$.
With this, the CST curve is completely characterized by the exponents $N_1$, $N_2$, the Bernstein coefficients $\mathbf{A} = \left(A_i\right)_{i=0}^{n}$ and $\zeta_T$.
\begin{figure}[ht]
    \centering
    \input{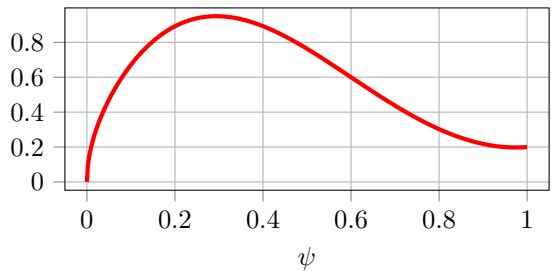}
    \caption{Example of an upper wing profile as a CST curve with parameters $N_1=0.5$, $N_2=1$, $\mathbf{A} = (2, 3, 2, 1)$ and $\zeta_T=0.2$.}
    \label{fig:profile_cst}
\end{figure}
Figure \ref{fig:profile_cst} shows a simple example of a CST curve for an upper wing profile.

\subsubsection{Guide Curves from Point Lists}
\label{ssub:guide_curves_from_point_lists}

Guide curves connect the profiles in order to create a curve network (cf.~\fref{fig:curve_network} and \Sref{sect:curve_network_interpolation}).
Each guide curve is created by B-spline interpolation of a set of guide curve points.
Since a guide curve always starts and ends at a profile, the first and the last guide curve points are attached to these profiles.
The position of the start points can be given either by a relative circumference of the profile or by pointing to the end point of a previous guide curve.
In the latter case, the continuity across the profile can be set.
This is described in \Sref{ssub:geometric_modeling_of_the_wing}.
The position of the end point is always set by a relative circumference of the profile.
The intermediate guide curve points are described by three local coordinates $(\alpha, \beta, \gamma)$.

In the following, we will describe how to construct a guide curve point in real space from the local coordinates (see~\fref{fig:guide_curve_points}).
As a first step, we draw a straight line from the start to the end point.
This line defines the first axis of the coordinate system. 
We move along this line from the start point ($\alpha=0$) towards the end point ($\alpha=1$).
Hereby, $\alpha$ is the normalized distance between start and end point.
From there, we move $\beta c(\alpha)$ towards a pre-defined direction.
Here $c(\alpha) = c_s (1 - \alpha) + c_e \alpha$ is the linear interpolation between the typical length of the start profile $c_s$ and the end profile $c_e$.
In the case of wing profiles, $c_s$ and $c_e$ are the cord lengths of the start and end profiles, respectively.
The pre-defined direction is usually the global $x$-axis for wing guide curves and the global $z$-axis for fuselage guide curves.
As a last step we move $\gamma c(\alpha)$ in the direction perpendicular to both previous directions.
\begin{figure}[ht]
    \centering
    \begin{tikzpicture}[scale=4]

\def\do{0.1}
\begin{scope}[shift={(-0.2,-0.2), scale=1}]
    \draw [<-,thick] (0.0,\do) node (zaxis) [above] {$z$} -- (0.0,0.0) node (origin)  {};
    \draw [->, thick] (0,0) -- ({\do*cos(30)},{\do*sin(30)}) node[right] {$y$};
    \draw [->,thick] (0.0,0.0) node (origin) {} -- (\do,0.0) node  (xaxis) [right] {$x$};
\end{scope}

\def\shiftx{0.7}
\def\shifty{0.5}
\def\startx{0.22652591}
\def\starty{0.06697582}
\def\endx{0.54128966}
\def\endy{0.0520147}
\def\Alpha{0.3}
\def\Beta{0.2}
\def\Gamma{0.2}

\definecolor{darkgreen}{rgb}{0,.6,0}
\coordinate (shift) at (\shiftx, \shifty);
\coordinate (endlocal) at (\endx, \endy);

\draw [black, thick] plot [smooth] file{./images/tikz/profile.dat};
\draw [-,dashed] (0.0, 0.0) -- node[below] {$c_s$}(1.0, 0.0) ;
\begin{scope}[shift={(\shiftx,\shifty), scale=1}]
    \draw [black, thick] plot [smooth] file{./images/tikz/profile.dat};
    \draw [-,dashed] (0.0, 0.0) -- node[below] {$c_e$}(1.0, 0.0) ;
\end{scope}

\node[draw,circle,minimum size=2pt,inner sep=0pt,outer sep=0pt,fill=black] (start) at (\startx, \starty) {};
\node[draw,circle,minimum size=2pt,inner sep=0pt,outer sep=0pt,fill=black] (end) at ($(endlocal) + (shift)$) {};

\draw [-,dashed] (start) node[below] {$\mathbf{s}$} -- (end) node[above] {$\mathbf{e}$};
\node[circle,minimum size=2pt,inner sep=0pt,outer sep=0pt,fill=blue] (x) at ($(start)!\Alpha!(end)$) {};
\draw [<->, thick, blue] (start) -- node[above=2pt] {$\alpha |\mathbf{d}|$} (x) ;

\coordinate (y) at ($(x) + (\Beta, 0)$);
\coordinate (yp) at ($(x) + (0.3, 0)$);
\draw [->, thick, black] (x) -- (yp) node[right=2pt] {$\mathbf{e}_x$} ;
\draw [<->, thick, darkgreen] (x) -- node[below=2pt] {$\beta c(\alpha)$} (y) ;

\coordinate (z) at ($(x) + (\Beta, 0) + (0, \Gamma)$ );
\coordinate (zp) at ($(x) + (0, 0.3)$);
\draw [->, thick, black] (x) -- (zp) node[above=2pt] {$\mathbf{e}_x \times \mathbf{d}$} ;
\draw [<->, thick, red] (y) -- node[right=2pt] {$\gamma c(\alpha)$} (z);
\node[circle,minimum size=5pt,inner sep=0pt,outer sep=0pt,fill=red, above] (guideCurvePoint) at (z) {};

\coordinate (legend) at (1.2, 0.0);
\node[right] (legend_d) at (legend) {$\mathbf{d} = \mathbf{e} - \mathbf{s}$};
\node[right] (legend_c) at ($(legend) - (0, 0.1)$) {$c(\alpha) = c_s (1 - \alpha) + c_e \alpha$};

\end{tikzpicture}
    \caption{Construction of a guide curve point in real space from local coordinates $(\alpha, \beta, \gamma) = (0.3, 0.2, 0.2)$ and two wing profiles.}
    \label{fig:guide_curve_points}
\end{figure}

\subsection{Wing}
\subsubsection{CPACS Parametrization of the Wing}

According to the CPACS definition, a wing is modeled from its (cross-)sec\-tions.
For a wing, at least two sections must be present, one for the root of the wing and one for the tip.

\begin{figure}[t]
\centering
\includegraphics[width=0.7\textwidth]{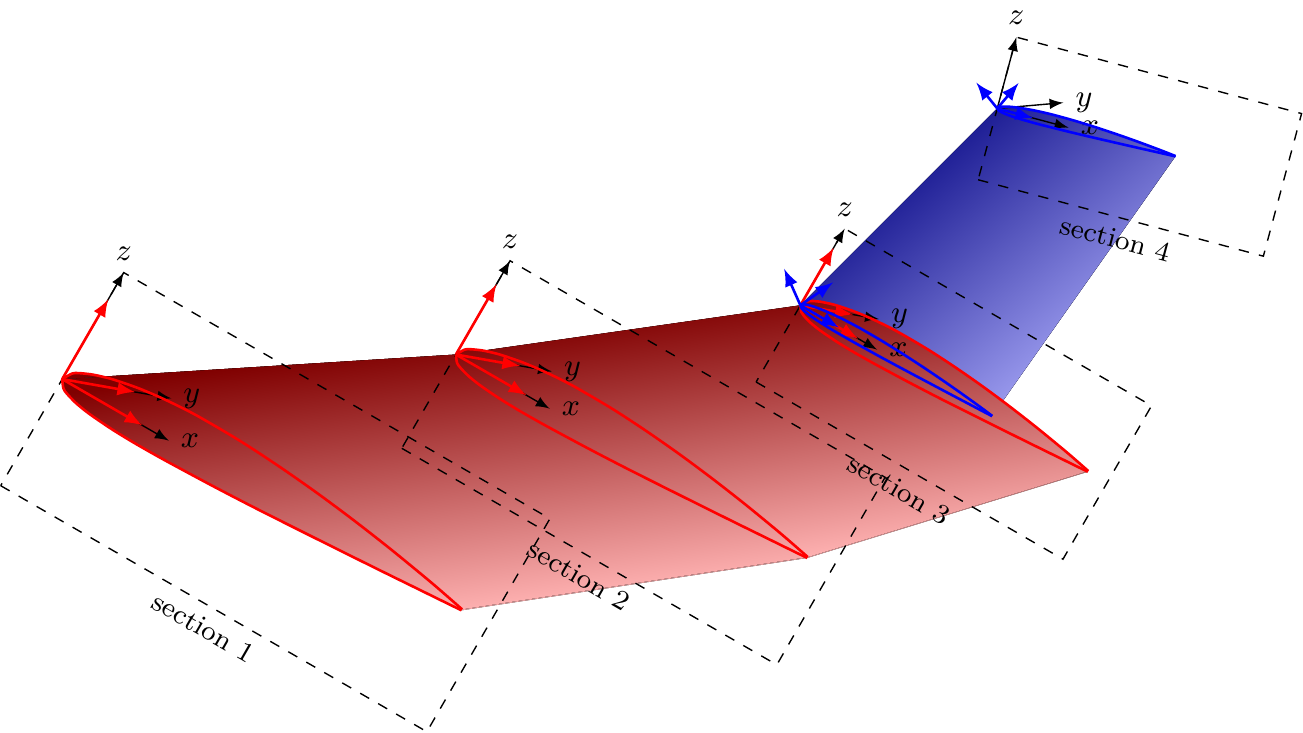}
\caption{Wing sections and elements according to the CPACS definition. Section 3 contains two elements with different rotations and scale. The wing therefore has a discontinuous shape at this section.}
\label{fig:cpacsdef_wing}
\end{figure}

A section is a coordinate system that is used to position airfoil curves in three-dimensional space, see \fref{fig:cpacsdef_wing}.
This coordinate system is defined using a transformation consisting of scaling in three dimensions, rotation around the $z$-, $y$- and $x$-axis, as well as a three-dimensional translation.
In addition to the transformations, sections can be translated relative to each other using a positioning vector. 
It consists of a sweep and dihedral angle, as well as a length for the offset between two sections.
The positioning vector of a section does not influence its rotation or scale.
The total translation of the section is the sum of the positioning vector in Cartesian coordinates and the translation prescribed in the section's transformation.

Within a section, several elements can be placed, where each element references one airfoil curve.
An element is again a coordinate system that is used to transform an airfoil within a section.
By placing two elements in one section, it is possible to define wings, whose cross section has a discontinuous jump in span-wise direction, see \fref{fig:cpacsdef_wing}.

A wing segment is the volumetric part of the wing that connects two elements from adjacent sections.
It is possible to use guide curves within each segment to influence the segment shape.
All segments must have the same number of guide curves and the guide curves of two adjacent segments must be connected.
Otherwise, the guide and profile curves would not constitute a valid curve network and the Gordon surface algorithm would fail. 

Finally, a component segment is a part of the wing that consists of several adjacent segments. 
Component segments are used to define the relative position of the internal wing structural elements and fuel tanks, control devices, and the wing fuselage attachment.

\subsubsection{Geometric Modeling of the Wing} 
\label{ssub:geometric_modeling_of_the_wing}
The wing profile curves described in section~\ref{ssub:profiles_from_point_lists} are elements of wing sections. 
They are transformed through the section's transformation and positioning vector, as well as the element's transformation itself.

Guide curves can be defined for the segments through guide curve points (cf.~\sref{ssub:guide_curves_from_point_lists}).
They are constructed as curves spanning the wing from its root to the tip.
Together with the wing profiles, they serve as input for the Gordon algorithm described in \sref{sect:gordon_surfaces}. 
The connected guide curves are interpolated piecewisely, depending on the prescribed continuity condition of the guide curve. 
Continuity conditions are optional, and they can include \emph{``C0''}, \emph{``C1 from previous''}, \emph{``C1 to previous''}, \emph{``C2 from previous''} and \emph{``C2 to previous''} according to the CPACS schema.

\begin{figure}[t]
\centering
\begin{tikzpicture}[scale=1.25]
\coordinate (TE_root) at ( 0, 0);
\coordinate (TE_s0)   at (-1, 1);
\coordinate (TE_s1)   at (-3, 1.5);
\coordinate (TE_tip)  at (-6, 1.3);
\coordinate (LE_tip)  at (-6, 2);
\coordinate (LE_s1)   at (-3, 2.5);
\coordinate (LE_s0)   at (-1, 2.5);
\coordinate (LE_root) at ( 0, 2);

\draw[line width=1pt, red] (TE_root) -- (LE_root);
\draw[line width=1pt, red] (TE_s0) -- (LE_s0);
\draw[line width=1pt, red] (TE_s1) -- (LE_s1);
\draw[line width=1pt, red] (TE_tip) -- (LE_tip);

\draw[line width=1pt, blue] (LE_tip) -- (LE_s1) node[midway,above,yshift=0.15cm]  {\scriptsize \emph{C1 to previous}} node[black,line width=0.5pt,draw,circle,midway,below,yshift=-0.1cm] {\tiny 3};
\draw[line width=1pt, blue] (LE_s1) to[out=9.4623, in=153.435] node[midway,above] {\scriptsize \emph{C1 from previous}} node[black,line width=0.5pt,draw,circle,midway,below,yshift=-0.1cm] {\tiny 2} (LE_s0);
\draw[line width=1pt, blue] (LE_s0) -- (LE_root) node[black,line width=0.5pt,draw,circle,midway,below,yshift=-0.1cm] {\tiny 1};

\draw[line width=1pt, blue] (TE_tip) -- (TE_s1) node[midway,above] {\scriptsize \emph{C0}} node[black,line width=0.5pt,draw,circle,midway,below,yshift=-0.1cm] {\tiny 6};
\draw[line width=1pt, blue] plot[smooth] coordinates {(TE_s1) (TE_s0) (TE_root)};

\node[black,line width=0.5pt,draw,circle,below,yshift=-0.1cm] at ($(TE_s1)!0.5!(TE_s0)$) {\tiny 5};
\node[black,line width=0.5pt,draw,circle,below,yshift=-0.1cm] at ($(TE_s0)!0.5!(TE_root)$) {\tiny 4};

\node at (TE_tip) {\Large $\bullet$};
\node at (TE_s1) {\Large $\bullet$};
\node at (TE_root) {\Large $\bullet$};
\node at (LE_tip) {\Large $\bullet$};
\node at (LE_s1) {\Large $\bullet$};
\node at (LE_s0) {\Large $\bullet$};
\node at (LE_root) {\Large $\bullet$};

\end{tikzpicture}
\caption{A wing modeled with guide curves. Guide curves 1, 4 and 5 have no continuity conditions prescribed. The (upper) leading edge is separated into three parts, where the middle part containing guide curve 2 must be interpolated after the parts containing guide curves 1 and 3. The trailing edge is broken into two parts. A ``$C^0$'' continuity condition for guide curve 6 is used to model a kink between the outer and middle segment.}
\label{fig:guidecurvecontinuity}
\end{figure}
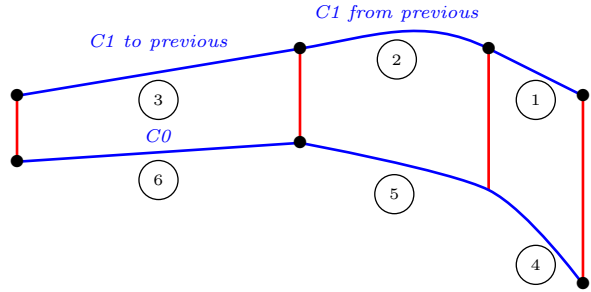

A connected guide curve is broken into parts at the prescribed continuity conditions, see \fref{fig:guidecurvecontinuity}. 
As a default, each part is interpolated smoothly, meaning a $C^2$ continuity is prescribed.
The parts depend on each other according to the \emph{``from previous''} or \emph{``to previous''} continuity conditions.
A \emph{``from previous''} conditions means, that the tangent at the beginning of the guide curve must be the same as the end tangent of the inner neighboring guide curve.
A \emph{``to previous''} condition implies, that the tangent at the beginning of the guide curve is prescribed onto the end of the inner neighboring guide curve.
This implies an order in which guide curve parts must be interpolated, so that the prescribed tangents are available.
TiGL uses a topological sorting algorithm based on Kahn's method \cite{kahn1962} to achieve this.
Note that TiGL only prescribes tangents  at the break points and not the curvature. 
Therefore, only $C^1$ continuity is guaranteed.

If there are no guide curves, the profile curves are skinned linearly, or optionally using a B-spline of degree up to three.
The resulting surface must be closed by side caps at the root and tip to make a solid, otherwise the wing geometry cannot be used in Boolean operations.
The modeling of the wing tip geometry is planned for the future.
\fref{fig:tigl_wing} shows a wing created from a CPACS file with TiGL.

\subsection{Flaps}
\begin{figure}[b]
\centering
\subcaptionbox{\label{fig:tigl_wing}}
{
  \includegraphics[width=0.45\textwidth]{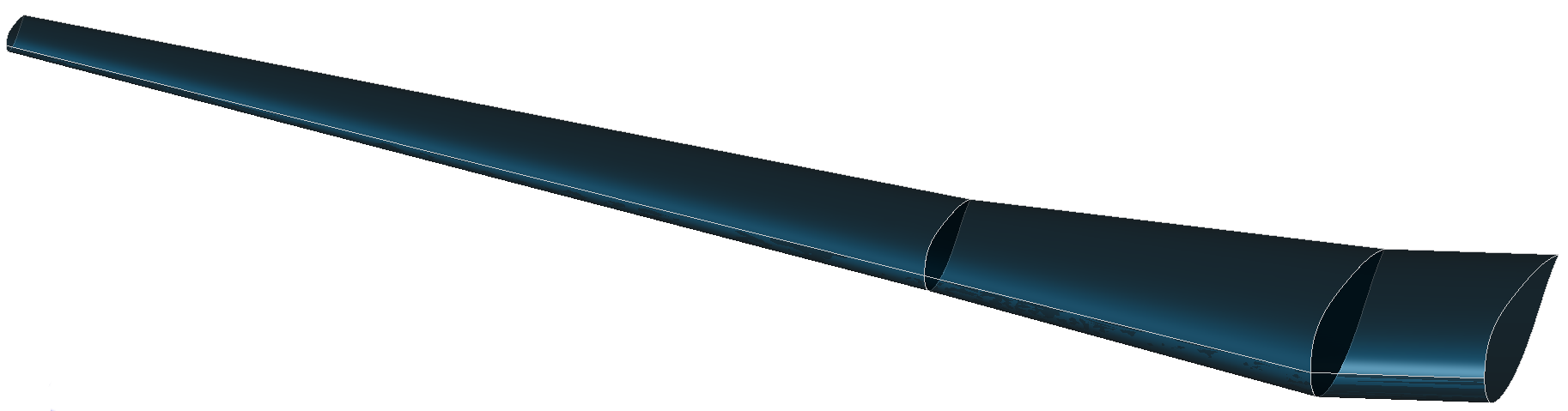}
}
\subcaptionbox{\label{fig:flaps}}
{
  \includegraphics[width=0.45\textwidth]{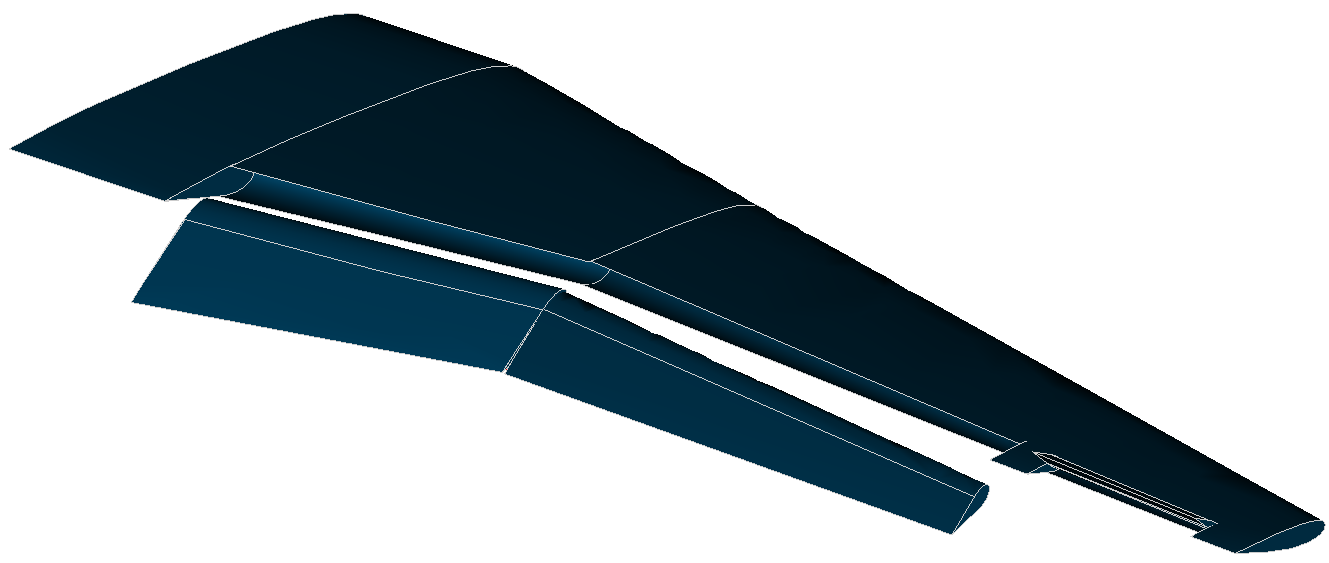}
}
\caption{A wing consisting of three segments, that was created with TiGL from a CPACS file.
The extended trailing edge devices of the same wing is shown in (B).
}
\end{figure}

Flight control surfaces such as ailerons, flaps, slats, spoilers and rudders can be modeled with CPACS and TiGL.
There are three categories: leading edge devices, trailing edge devices and spoilers.
\fref{fig:flaps} shows extended trailing edge devices that were modeled using TiGL.
The devices can have an internal structure, that is similar to the definition of the wing structure, see \Sref{sect:structure}.

The outer shape of a control surface is defined by defining four points in the local $(\eta,\xi)$-coordinates of the component segment. 
These points roughly describe the position and shape of the control device as well as the wing cutouts.
Alternatively, the exact shape of the flap can be described using profile curves.
In addition to the exact control surface shape, the shape of the wing cutout can be described more precisely by defining the cutout limits independently of the flap shape and separately for the upper and lower skin of the wing. 
In this case, the upper and lower cutout must be closed with a profile curve on the inner and outer side of the flap.

Alongside the description of the actuators of the control surfaces, the path along which a control surface moves when it is extended can be given. 
For this, an inner hinge point and an outer hinge point must be provided in order to define the connection of the flap to the actuators. 
All possible positions of the flap are described by giving steps along a path from a minimum deflection to a maximum deflection value.
A step along this path includes a translation of the control device together with a rotation around the axis defined by the two hinge points.

Using TiGL's API, the flaps can be deflected and the resulting geometry can be exported for further processing.
A console in TiGL Viewer allows the interactive deflection of the control devices for more control.

\subsection{Fuselage}
\begin{figure}[h]
\centering
\includegraphics[width=0.7\textwidth]{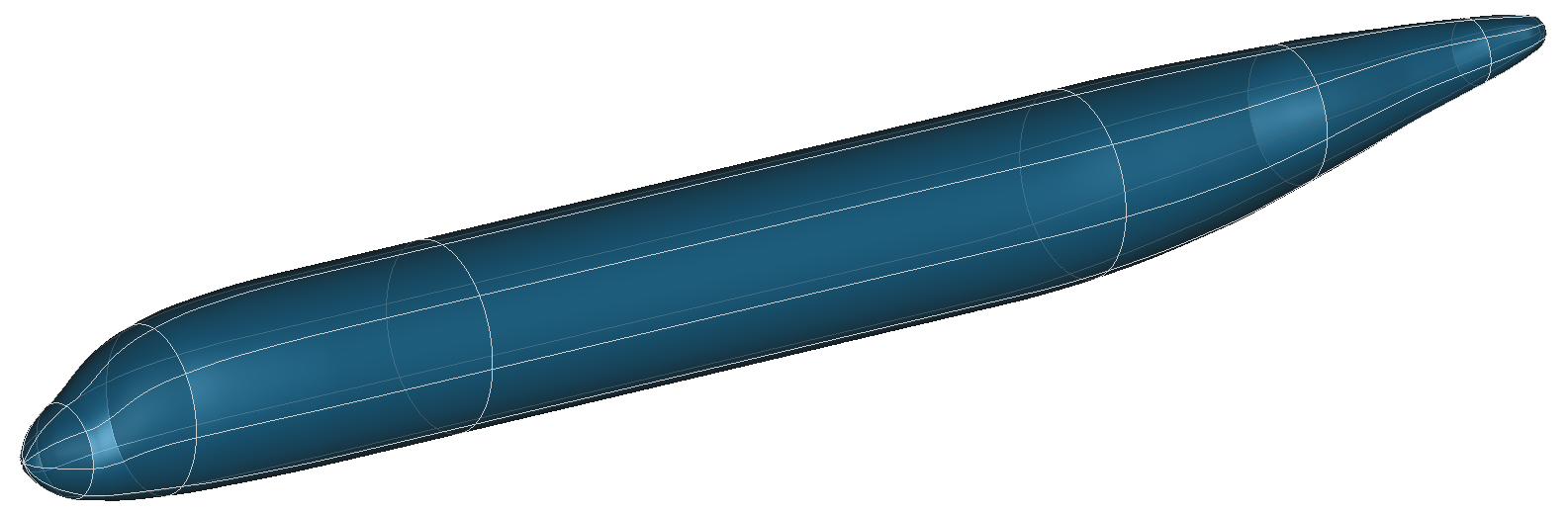}
\caption{A fuselage built from eight profile curves and eight guide curves.}
\label{fig:fuselage}
\end{figure}

Fuselages are created in TiGL similarly to wings: A fuselage is comprised of segments, which contain sections.
Each section can contain one or more elements of fuselage profiles.
The sections can be connected via guide curves.
In order to create a solid fuselage, the profile curves and the connected guide curves with prescribed continuity conditions are transformed to global coordinates in TiGL.
Together, they form the curve network that is used as an input for the Gordon Surface algorithm. 
Currently, the front and back of the fuselage are closed by side caps to create solids.
The modeling of fuselage noses and rears are planned for the near future.
\fref{fig:fuselage} shows a fuselage that was created with TiGL.

\subsection{Wing and Fuselage Structure} \label{sect:structure}
\subsubsection{Wing structure geometry}
\label{sec:wing_structure}
\begin{figure}[b!]
	\centering
	\subcaptionbox{\label{fig:spardefinition}}
	{\includegraphics[width=0.33\textwidth]{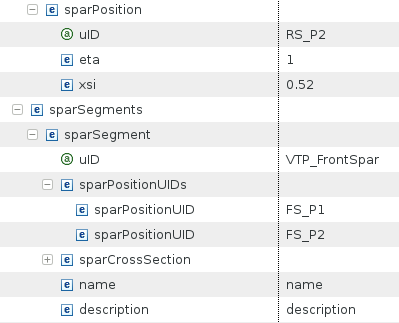}}
	\qquad
	\subcaptionbox{\label{fig:sparConstruction}}{	
\begin{tikzpicture}
    \node[anchor=south west,inner sep=0] (image) at (0,0) {\includegraphics[width=0.42\textwidth]{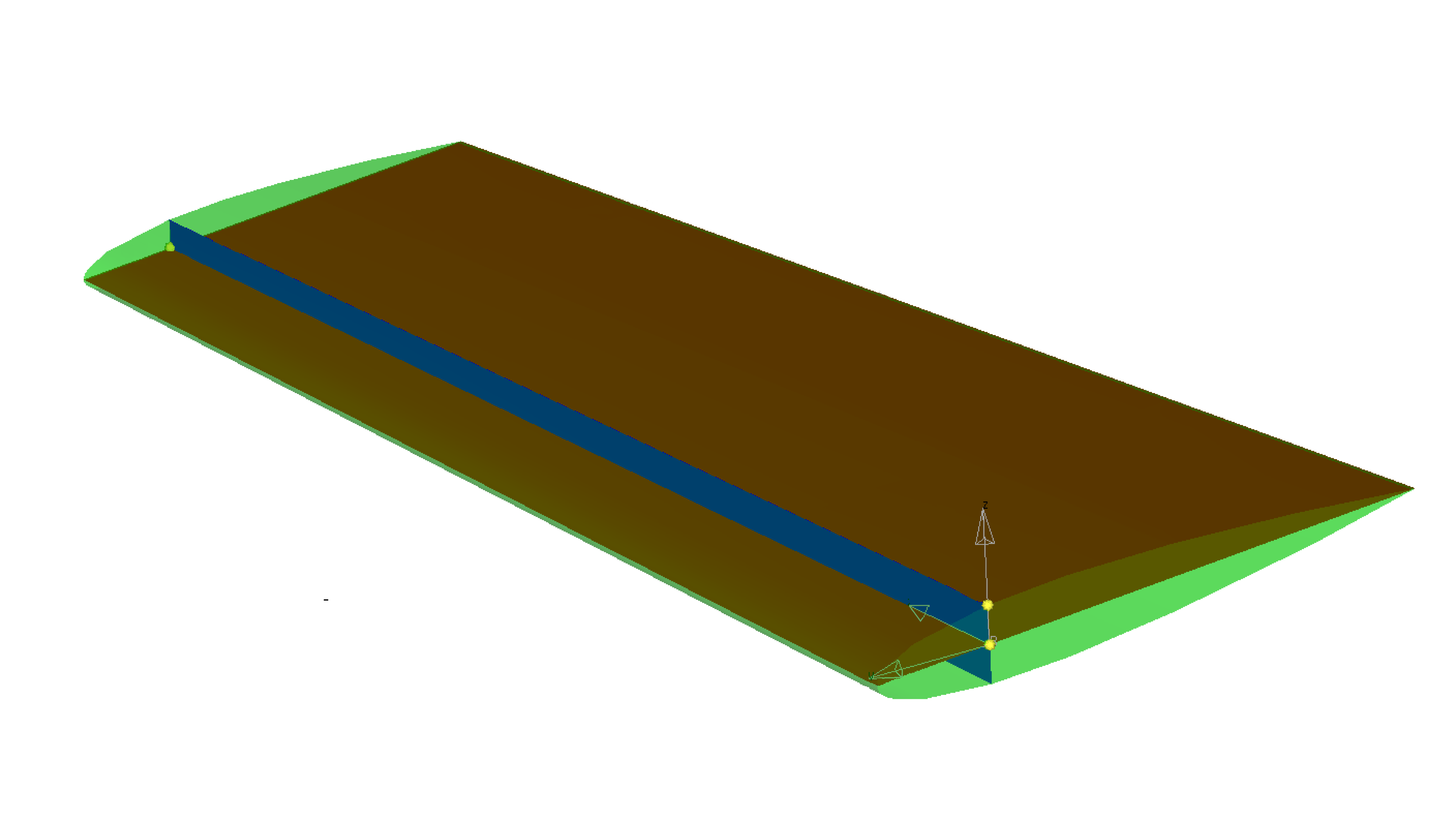}};
    \begin{scope}[x={(image.south east)},y={(image.north west)}]
        \draw  (0.7, 0.8) node (aa) {\scriptsize Construction points};
        \draw  (0.85, 0.65) node (bb) {\scriptsize Midplane};
        \draw (0.1, 0.4) node (ff) {\scriptsize Wing loft};
        \draw (0.2, 0.21) node (dd){\scriptsize Spar face};
        \node[circle,fill=black,inner sep=0pt,minimum size=2pt] (a) at (0.678,0.244) {};
        \node[circle,fill=black,inner sep=0pt,minimum size=2pt] (b) at (0.13,0.7) {};
        \node[circle,fill=black,inner sep=0pt,minimum size=2pt] (c) at (0.75,0.4) {};
        \node[circle,fill=black,inner sep=0pt,minimum size=2pt] (d) at (0.49,0.4) {};
        \node[circle,fill=black,inner sep=0pt,minimum size=2pt] (e) at (0.62,0.187) {};
        \node[circle,fill=black,inner sep=0pt,minimum size=2pt] (f) at (0.09,0.69) {};
        \draw (0.25, 0.01) node (ee) {\scriptsize Spar plane normal axis};
        \draw  (aa) -- (a) ;
        \draw  (aa) -- (b) ;
        \draw  (bb) -- (c) ;
        \draw  (dd) -- (d) ;
        \draw  (ee) -- (e) ;
        \draw  (ff) -- (f) ;
    \end{scope}
\end{tikzpicture}
	}
	\caption{CPACS Definition (A) and construction (B) of the wing spars.}
\end{figure}
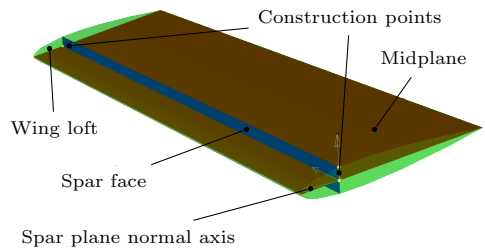

The structural definition of the wing is being developed by \emph{Airbus Defence and Space}
since $2012$ \cite{MPD13} and was cantributed back to the TiGL source code in $2015$.
The foundation is the wing component segment, consisting mainly of the upper and lower wing shell,
wing stringers, spars and ribs.
Currently only the geometry generation of the ribs and spars is supported by TiGL.

The spar definition, shown in \fref{fig:spardefinition}, is realized with spar positions and segments.
A positioning can be defined by ($\eta$, $\xi$) coordinates in the
relative space of the component segment or with an unique identifier (UID) referring to a section-element and a $\xi$ value.
A spar segment has to consist of two or more spar positions.

The rib geometries are constructed as a second step,
because their definition can be spar dependent.
There are two different rib positioning schemes, the common and the explicit one.
The common rib positioning is defined by two $\eta$ values, one for the first and one for the last rib.
As for the spar positions,
these values can also be replaced by section-element UIDs,
to place ribs directly at a section border.
Additionally, the number of ribs is defined in the CPACS schema, and so the remaining ribs are
placed equally distributed between the start and end rib.
The chord-wise borders of the ribs can be the leading or the trailing edge of the wing
or any spar that intersects the rib plane.

\begin{figure}[t]
\centering
\subcaptionbox{Three different rib sets.\label{fig:ribConstruction}}
{
  \includegraphics[width=0.37\textwidth]{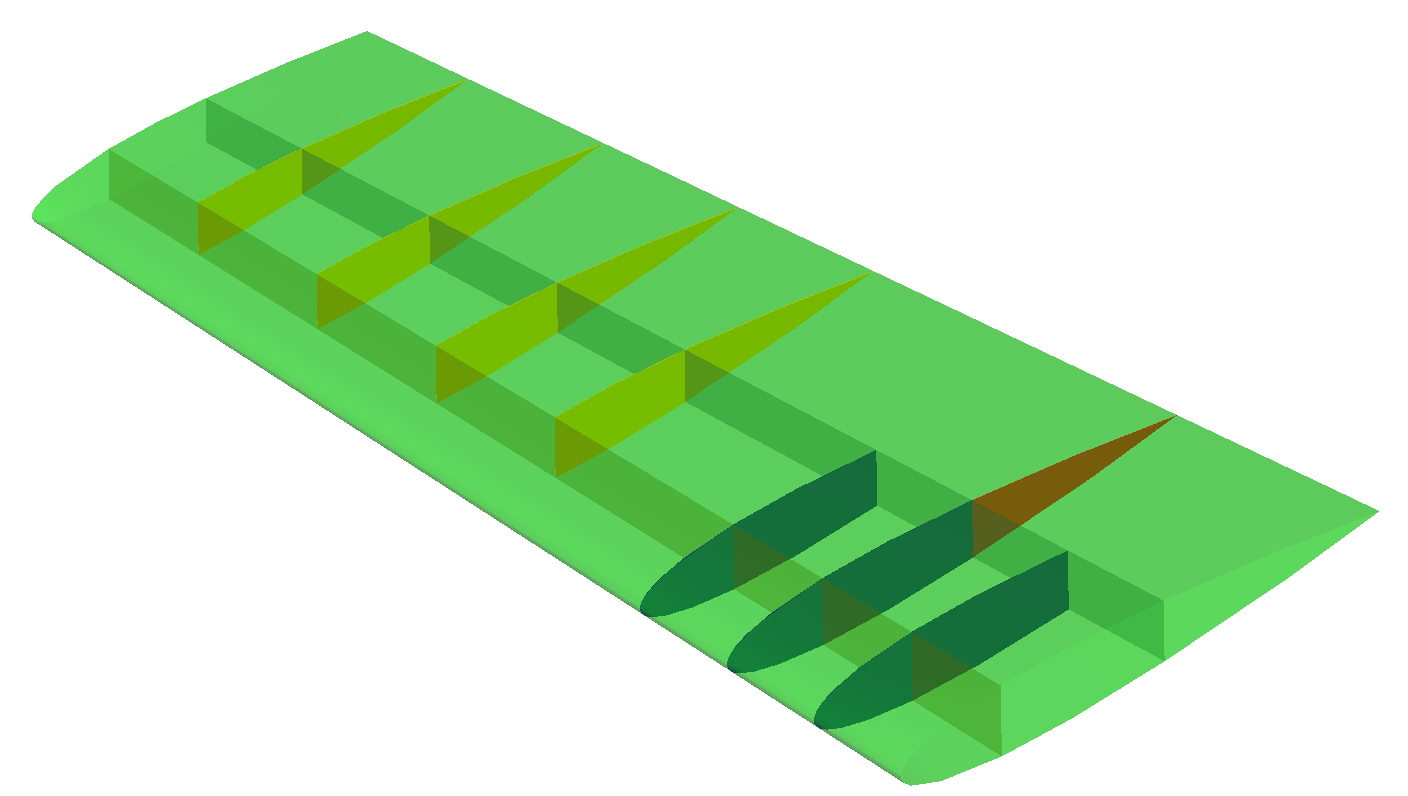}
}
\qquad
\subcaptionbox{One rib with three explicitly defined rib faces.\label{fig:ribExplicitConstruction}}
{
	\includegraphics[width=0.37\textwidth]{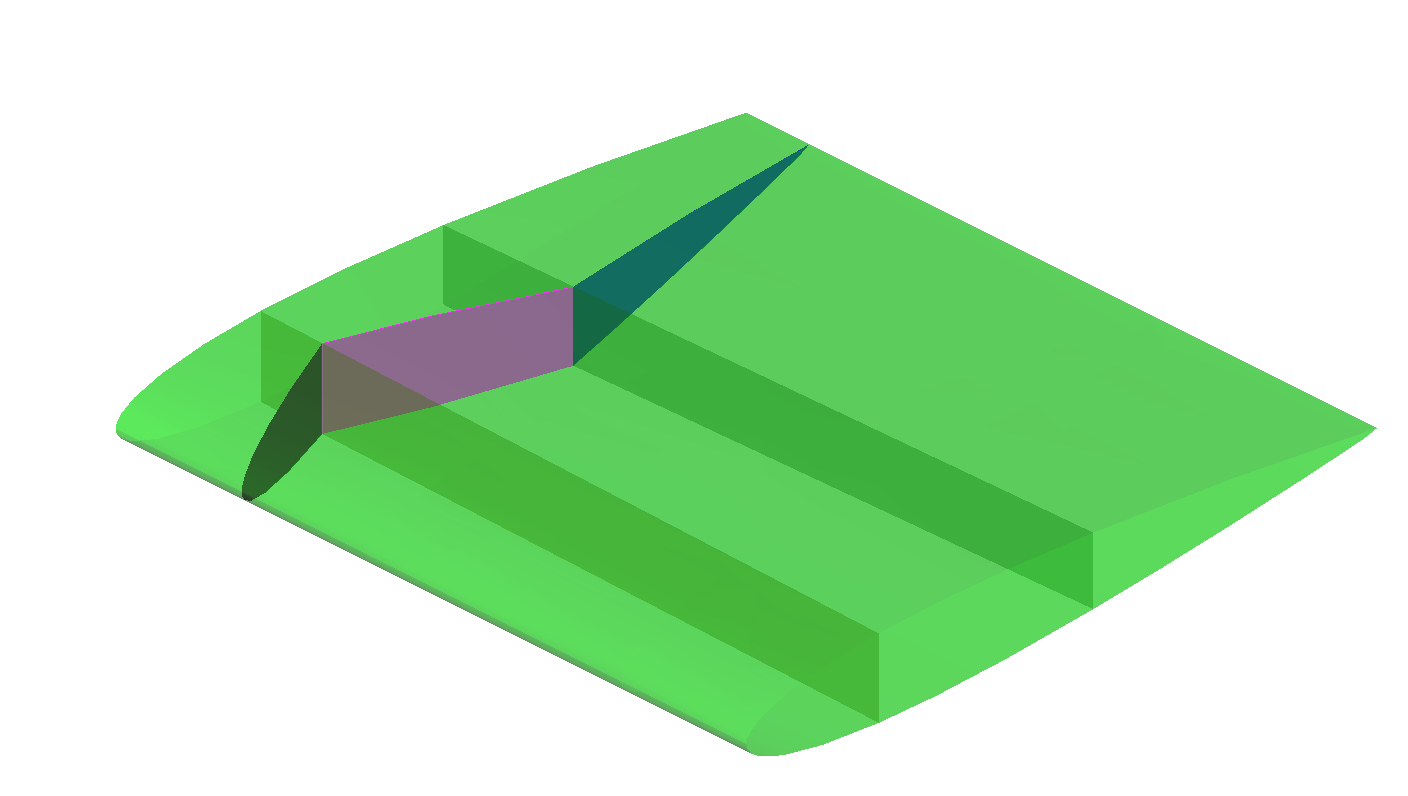}
}
\caption{The different options for the wing ribs definition.} 
\end{figure}

The main issue with this CPACS definition is, that it is not possible
to define three or more ribs with a dedicated chord-wise connection.
This is due to the common rib definition, which uses one span-wise position in combination with an angle.
This allows to generate either an exact starting point or an exact ending point.
To encounter this, the explicit rib definition with an exact chord-wise start and end position, was introduced.
A major drawback of this method is, that every single rib face has to be defined and no distribution can be given.
The described difference is visualized in \fref{fig:ribConstruction} and \fref{fig:ribExplicitConstruction}.

\subsubsection{Fuselage structure geometry}
\label{sec:fuselage_structure}

\begin{figure}[b]
		\centering
		\begin{tikzpicture}
    \node[anchor=south west,inner sep=0] (image) at (0,0) {\includegraphics[height=5cm]{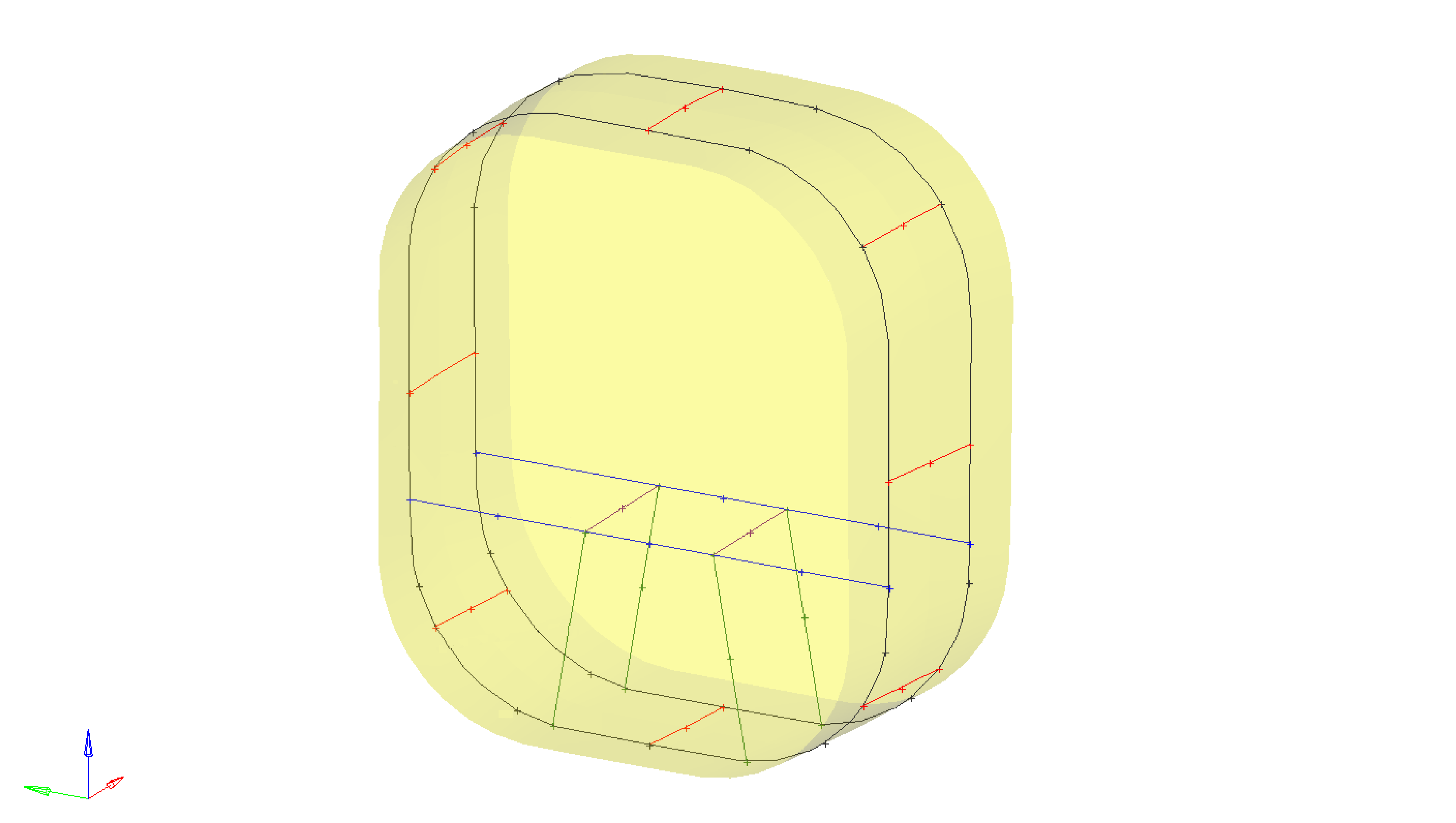}};
    \begin{scope}[x={(image.south east)},y={(image.north west)}]
        \draw  (0.10, 0.74) node (aa) {\scriptsize Frames};
        \draw  (0.10, 0.61) node (bb) {\scriptsize Stringer};
        \draw (0.10, 0.48) node (cc) {\scriptsize Cross Beams};
        \draw (0.09, 0.343) node (dd){\scriptsize Long Floor Beams};
        \draw (0.10, 0.23) node (ee) {\scriptsize Cross Beam Struts};
        \node[circle,fill=black,inner sep=0pt,minimum size=2pt] (a) at (0.282,0.72) {};
        \node[circle,fill=black,inner sep=0pt,minimum size=2pt] (b) at (0.303,0.546) {};
        \node[circle,fill=black,inner sep=0pt,minimum size=2pt] (c) at (0.388,0.43) {};
        \node[circle,fill=black,inner sep=0pt,minimum size=2pt] (d) at (0.425,0.377) {};
        \node[circle,fill=black,inner sep=0pt,minimum size=2pt] (e) at (0.39,0.24) {};

        \draw  (aa) -- (a) ;
        \draw  (bb) -- (b) ;
        \draw  (cc) -- (c) ;
        \draw  (dd) -- (d) ;
        \draw  (ee) -- (e) ;
        \draw  (0.10, 0.065) node {\tiny x};
        \draw  (0.005, 0.05) node {\tiny y};
        \draw  (0.062, 0.14) node {\tiny z};
    \end{scope}
\end{tikzpicture}
\caption{Fuselage segment with one-dimensional structure.}
\label{fig:fuseSegment}
\end{figure}
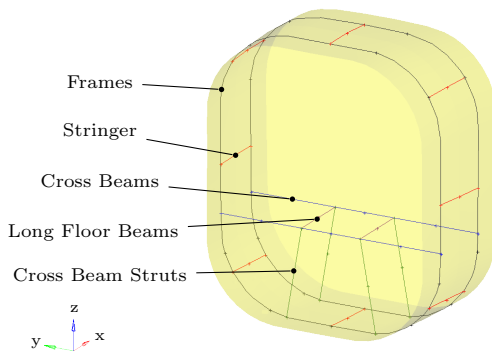

The structural definition of the fuselage is also developed by \emph{Airbus Defence and Space} and was contributed to TiGL in $2018$.
The CPACS definition of the fuselage structure differs completely from the one of the wing. Unfortunately, it is based on absolute coordinates, which leads to problems with the paradigm of parametric modeling. This was solved with an internal normalization of the absolute coordinate values in TiGL. Global $x$ values are normalized with the overall length of the fuselage. The $y$ and $z$ values are normalized with a bounding box, containing a curve of the fuselage loft at a global $x$ position. This solution enables the automatic adaption of the fuselage structure, if the fuselage loft is changed. When a CPACS export is requested, the absolute values are calculated in the reverse way and exported in absolute numbers.

The structural entities of the fuselage currently supported by TiGL, are: 
skin cells, fuselage frames, fuselage stringer, pressure bulkheads, fuselage doors,
cross beams, cross beam struts, and long floor beams.
Some of these entities are shown in \fref{fig:fuseSegment}.

\subsection{Engine Nacelles and Pylons}
\begin{figure}[ht]
\centering
\begin{subfigure}[t]{0.45\textwidth}
\includegraphics[width=\textwidth]{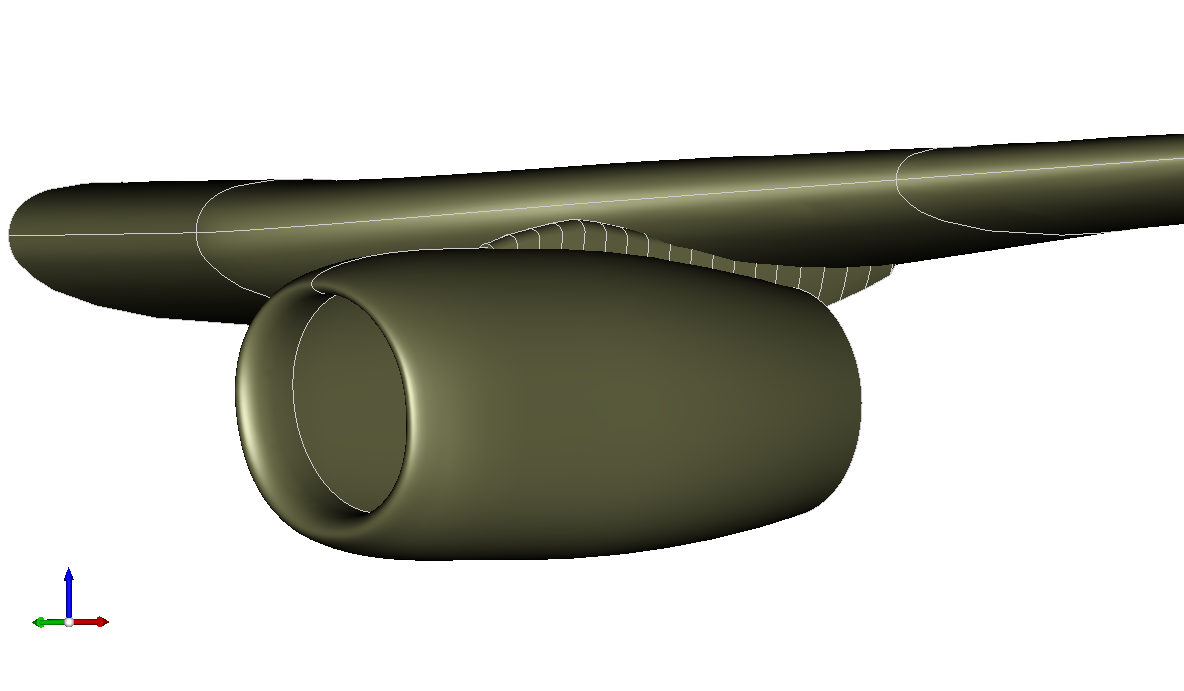}
\caption{A wing with engine nacelle and pylon. Here, the engine is loaded from an STEP file as a \emph{generic geometry component}.}
\label{fig:pylon1}
\end{subfigure}
\qquad
\begin{subfigure}[t]{0.33\textwidth}
\includegraphics[width=\textwidth, trim=0 -300 0 0, clip]{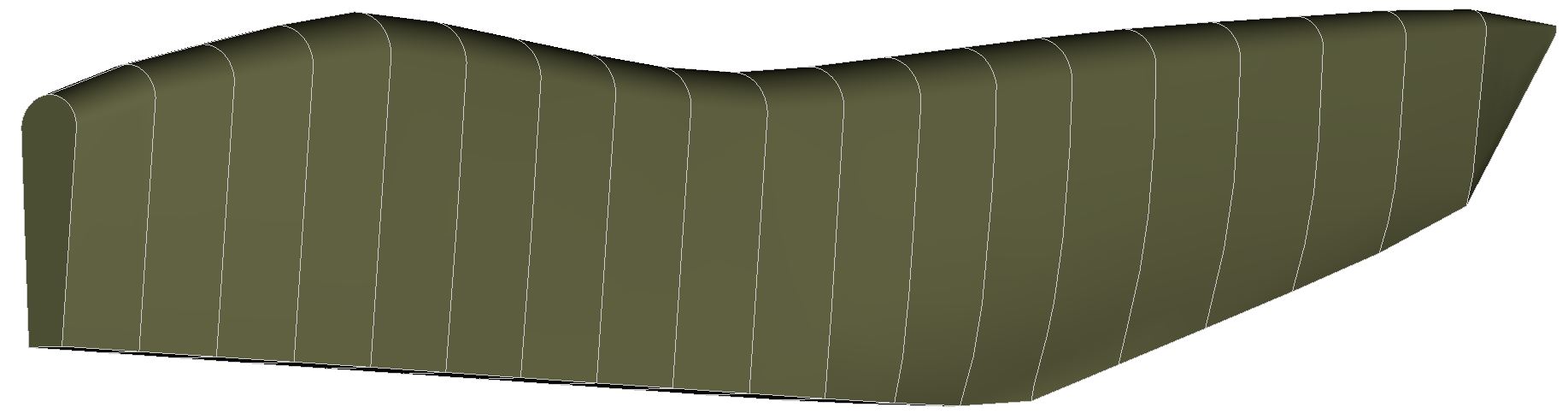}
\caption{A pylon is generated from profile curves skinned in chord-wise direction.}
\label{fig:pylon2}
\end{subfigure}
\caption{Modeling of engine nacelles and pylons.}
\label{fig:pylonnacelle}
\end{figure}

Engines are connected to the wing by pylons (see \Fref{fig:pylonnacelle}). 
Engine nacelles and pylons are not yet implemented in TiGL, but will be in the near future. 
The definition of the pylon and engine geometry is currently undergoing changes in CPACS as well. The definition for the pylon geometry as described here is already included in the latest CPACS release, while the engine nacelle definition will be updated soon.
The CPACS definition of pylons resembles that of a wing. Profile curves define the span-wise cross-section of a pylon and the curves are skinned in chord-wise direction (see \fref{fig:pylon2}).
In accordance with the CPACS definition, the outer geometry of the engine nacelles will be modeled from profile and guide curves (see \fref{fig:curve_network}).
Two-dimensional profile curves define the radial sections of the engine nacelle in flow direction. 
The profile curves are placed around the engine's symmetry axis using cylindrical coordinates, i.e. by prescribing an angle and a radius.
If only one profile is given, the resulting engine nacelle will be rotationally symmetric.
Otherwise, the profiles will be connected using closed guide curves.

To guarantee that the inner geometry of the engine nacelle is perfectly rotationally symmetric, a curve for the inner shape can be defined with an offset from the engines symmetry axis.
This curve is used to generate a rotation surface which is then blended with the -- not necessarily rotationally symmetric -- outer nacelle surface in a transition zone.

\section{Summary and Outlook} \label{sect:conclusion}
\begin{figure}[t]
\centering
\subcaptionbox{DLR-D150}
{
  \includegraphics[width=0.45\textwidth]{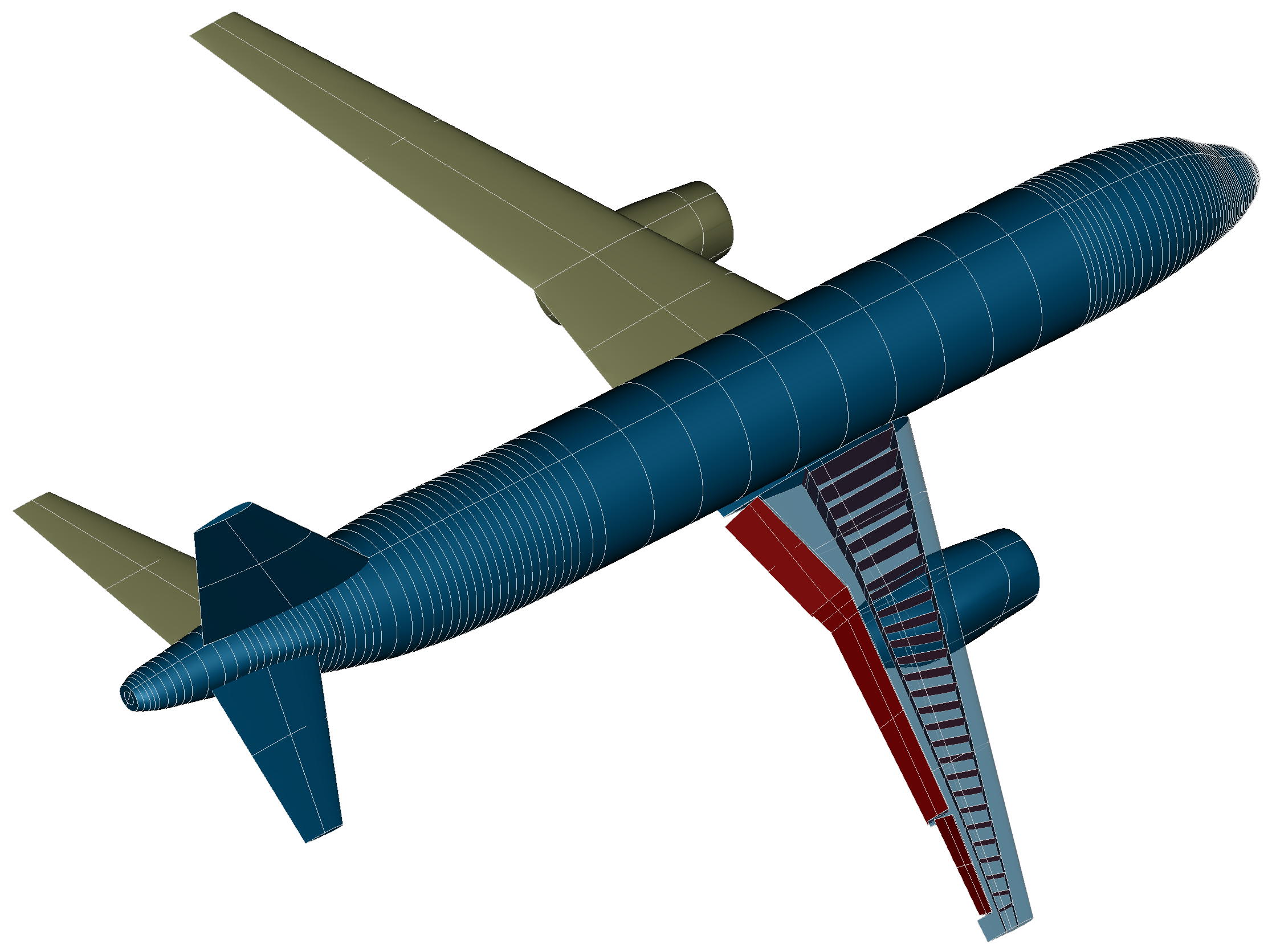}
}
\qquad
\subcaptionbox{NATO AVT 251 UCAV –- MULDICON}
{
	\includegraphics[width=0.45\textwidth]{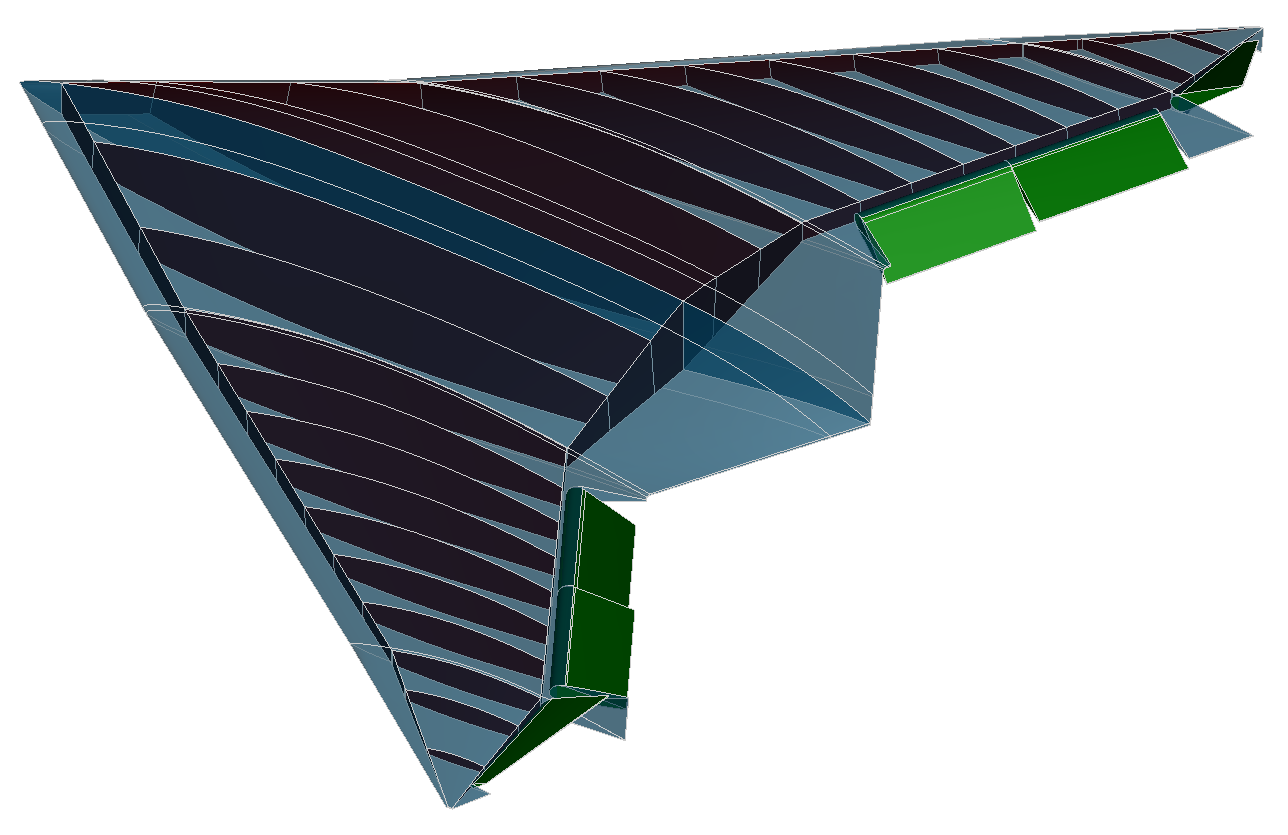}
}
\qquad
\subcaptionbox{Simple Helicopter}
{
	\includegraphics[width=0.45\textwidth]{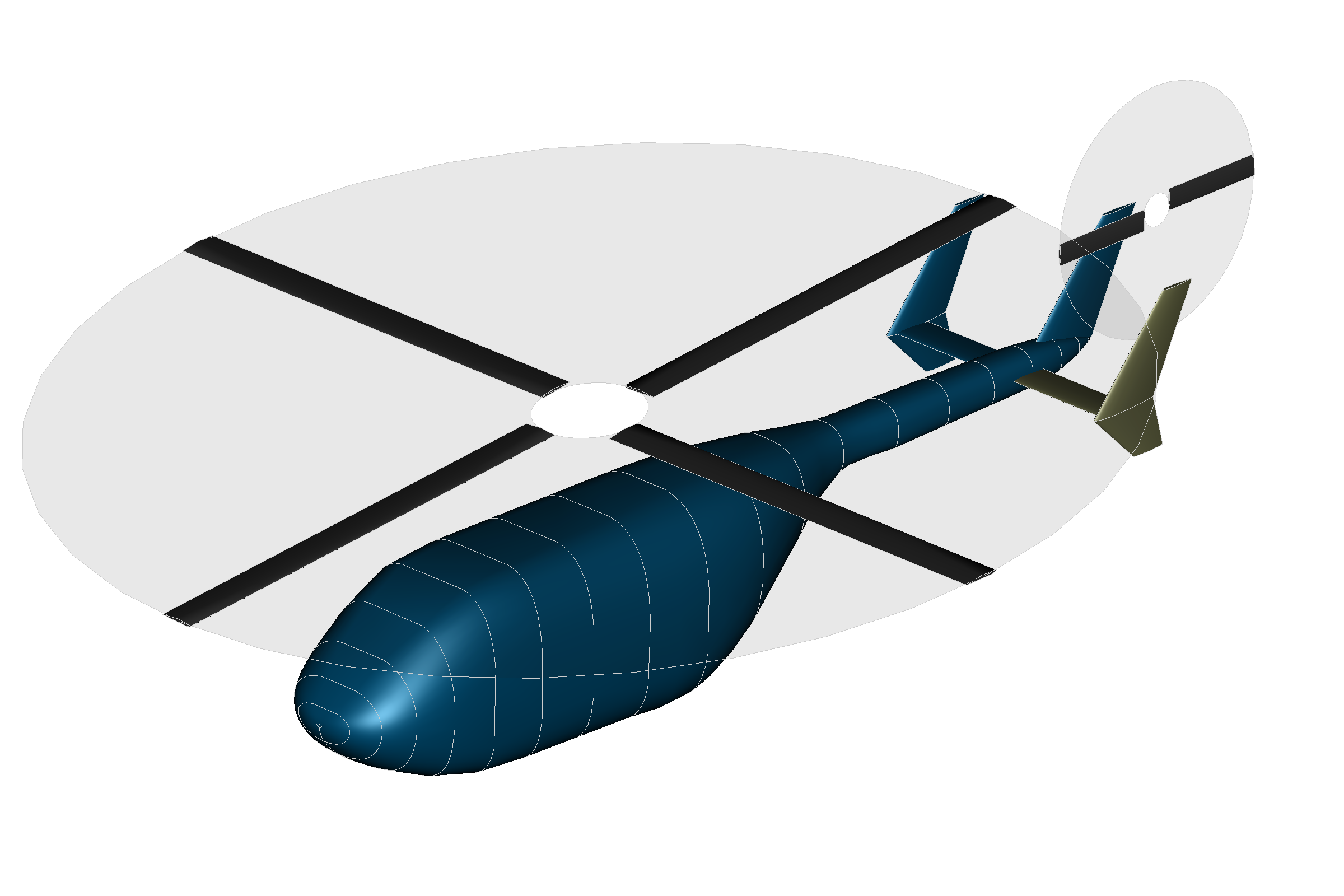}
}
\subcaptionbox{Ariane 5 Rocket}
{
	\includegraphics[width=0.45\textwidth]{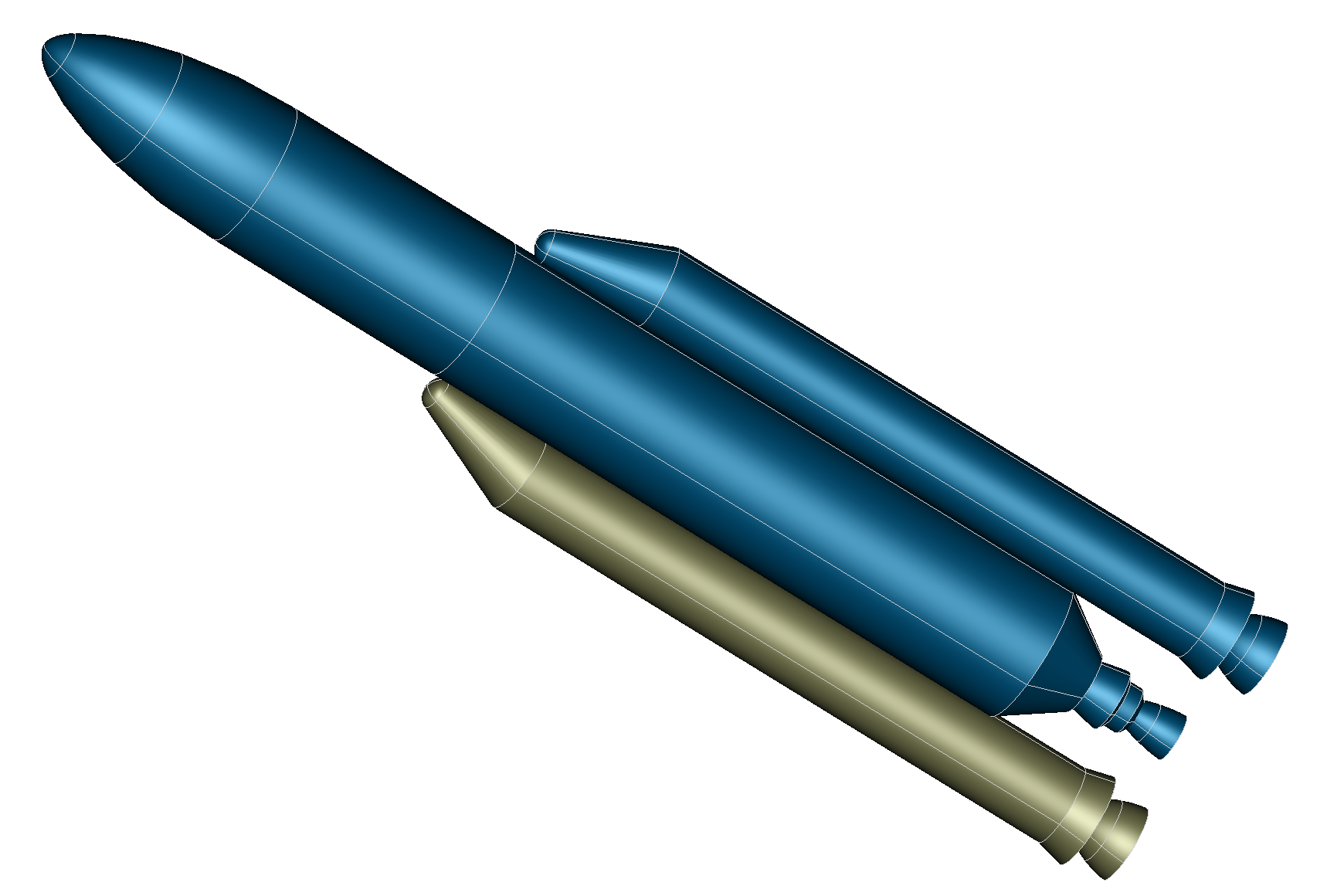}
}
\qquad
\subcaptionbox{Blended Wing Body}
{
	\includegraphics[width=0.45\textwidth]{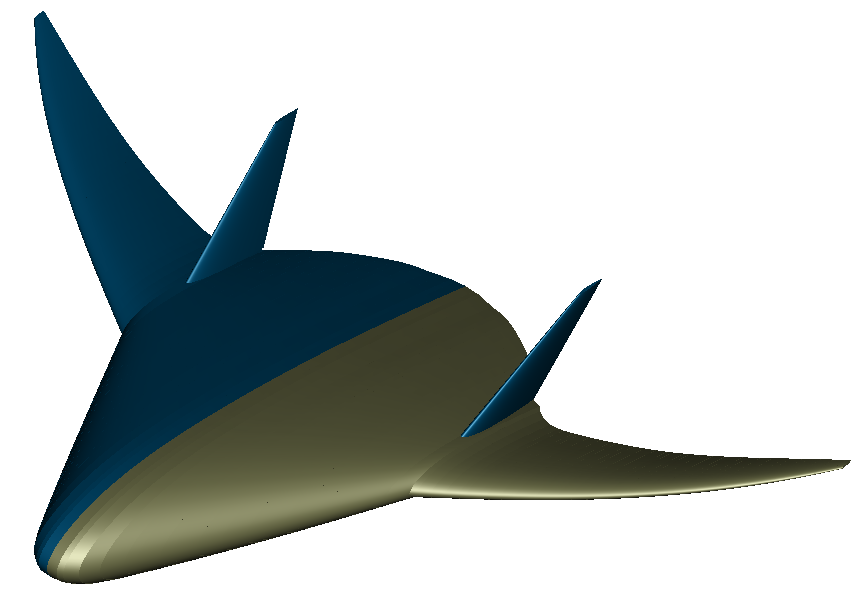}
}
\caption{Different design configurations created with TiGL.}
\label{fig:examples}
\end{figure}

This paper presented the software TiGL, which is a parametric geometry
generator for aircraft-like configurations.
It can be used in the aircraft design optimization process, 
by changing the CPACS design variables of the aircraft and
regenerating the geometry using TiGL.
TiGL models the major parts of an aircraft, including wings,
fuselages, control surface devices, the inner aircraft structure,
nacelles and pylons.
It is primarily focused on parametric aircraft configurations
in the CPACS format, which is getting more and more traction in the
aircraft design community.
Using TiGL and CPACS, it is possible to model a broad range
of different configurations.
\Fref{fig:examples} illustrates five different example configurations,
all defined in the CPACS format and modeled with TiGL.

Moreover, since TiGL is modular, its core modules can also be used completely disconnected from CPACS.
For example it is possible to use only TiGL's geometry module for general modeling.
One of the most important features of the geometry module is the implementation of the B-spline based Gordon surface algorithm.
To the authors knowledge, this algorithm almost never occurs in freely available software. 
Since this algorithm allows for high precision surface modeling,
TiGL is also suitable for high fidelity analysis.

TiGL is already used by the aircraft community outside the Ger\-man Aero\-space Center (DLR).
For example \emph{Airbus Defence and Space} developed the \emph{DESCARTES} analysis software \cite{decartes}
based on TiGL and improved TiGL during their development.
A graphical CPACS editor \cite{cpacs_editor} is currently being developed
by \emph{CFS Engineering} and is based on the TiGL Viewer.
The aircraft tool suite \emph{JPAD} \cite{jpad} is using TiGL
to integrate CPACS support into their software.

The development of TiGL will continue.
In the near future, we will finish our implementation
on nacelles and pylons.
Afterwards, belly fairings, an improved modeling of the wing tips
and winglets will follow.
We are currently working on the automatic mesh generation 
for low- and mid-fidelity CFD. 
Therefore, it is planned to integrate the Salome Mesh module \cite{salome},
which is based on Netgen \cite{netgen}, such that
surface and volumetric meshes can be generated via a call to TiGL's API.
To improve the support of gradient based MDO, it will be investigated whether
an adjoint code or automatic differentiation
of the geometry kernel is possible.

\subsection*{Acknowledgement}
TiGL has been developed for several years now.
During this time TiGL has been developed and improved by many of our colleagues.
In particular we would like to thank Markus Litz, who laid the foundation for TiGL.
Special thanks go to Bernhard Gruber and Roland Landertshammer from RISC Software for their work on the software development and Merlin Pelz from DLR for the Gordon surface implementation. Finally, many thanks to our colleagues Jonas Jepsen, Philipp Kunze, Sebastian Deinert, Mark Geiger, Volker Poddey, Konstantin Rusch, and Paul Putin for their contributions to TiGL.

\bibliographystyle{ieeetr}
\bibliography{references}

\end{document}